\documentclass[11pt]{article}
\usepackage{graphicx,float,bigdelim,bigstrut,multirow,pdflscape,amsthm, amsmath, amssymb,hyperref}
\usepackage[noend]{algorithmic}
\newtheorem{myprop}{Proposition}
\title{The price impact of order book events}
\author{Rama Cont, Arseniy Kukanov  and Sasha Stoikov}
\date{March 2011}
\begin{document}
\maketitle
\thispagestyle{empty}
\begin{abstract}
We study the price impact of order book events - limit orders, market orders and cancelations - using the NYSE TAQ data for 50 U.S. stocks. We show that, over short time intervals, price changes are mainly driven by the {\it order flow imbalance}, defined as the imbalance between supply and demand at the best bid and ask prices.
  Our study reveals a linear relation between order flow imbalance and price changes, with a slope inversely proportional to the market depth. These results are shown to be robust to seasonality effects, and stable across time scales and across stocks. We argue that this linear price impact model, together with a scaling argument, implies the empirically observed ``square-root" relation between price changes and trading volume. However, the relation between price changes and trade volume is found to be noisy and less robust than the one based on order flow imbalance.
\end{abstract}
\tableofcontents
\newpage
\section{Introduction}
The availability of high-frequency records of trades and quotes has stimulated  an extensive empirical and theoretical literature on the relation between order flow, liquidity and price movements in order-driven markets. A particularly important issue for applications   is the impact of orders on prices: the  optimal liquidation of a large block of shares, given a fixed time horizon,  crucially involves assumptions on price impact (see Bertsimas and Lo \cite{bertsimas98}, Almgren and Chriss \cite{almgren00}, Obizhaeva and Wang \cite{obizhaeva05}).

 Various models for price impact have been proposed in the literature but there is little agreement on how to model it \cite{bouchaud09-enc}.
 In the empirical literature, price impact has been described by various authors as linear, non-linear, square root, virtual, mechanical, temporary, instantaneous, permanent or transient. The only consensus seems to be the intuitive notion that imbalance between supply and demand moves prices.

The empirical literature on  price impact  has primarily focused on trades. One approach is to study the   impact of ``parent orders"  gradually executed  over time  using proprietary data (see Engle et. al \cite{engle06}, Almgren et. al \cite{almgren05}). Alternatively, empirical studies on  public data \cite{evans02,gabaix03,hasbrouck91,keim96,kempf99,torre97,plerou02,potters03} have  investigated  the relation between the direction and sizes of trades  and price changes and typically conclude that the price impact of trades is an increasing, concave (``square root") function of their size. This focus on trades leaves out the information in quotes,  which provide a more detailed picture of price formation  \cite{engle03}, and raises a natural question: is volume of trades truly the best explanatory variable for price movements in markets where many quote events can happen between two trades?

Understanding the price impact of orders is also important from a theoretical perspective, in the context of optimal order execution. Huberman and Stanzl \cite{huberman04} show that there are arbitrage opportunities if the effect of trades on prices is permanent and the impact is non-linear;  Gatheral \cite{gatheral10} extends this analysis
 by showing that if the price impact function is non-linear, impact needs to decay in a particular way to exclude arbitrage. Bouchaud et al. \cite{bouchaud04} associated the decay of price impact of trades  with limit orders, arguing that there is
a ``delicate interplay between two opposite tendencies: strongly correlated market orders that lead to super-diffusion (or persistence), and mean reverting limit orders that lead to sub-diffusion (or anti-persistence)". This insight implies that looking solely at trades, without including the effect of limit orders amounts to ignoring an important part of the price formation mechanism.

There is ample evidence that limit orders play an important role in determining price dynamics. Arriving limit orders significantly reduce the impact of trades \cite{weber05} and the concave shape of the price impact function changes depending on the contemporaneous limit order arrivals \cite{stephens09}. 
The outstanding limit orders (also known as market depth) significantly affect the impact of an individual trade (\cite{knez96}) and low depth is associated with large price changes \cite{weber06,farmer04}. Hasbrouck and Seppi \cite{hasbrouck01} use depth as one of the factors that determine price impact. The emphasis in these studies remains, however, on trades and there are few empirical studies that focus on limit orders from the outset. Notable exceptions are Engle \& Lunde \cite{engle03}, Hautsch and Huang \cite{hautsch09} who perform an impulse-response analysis of limit and market orders, Hopman \cite{hopman07} who analyzes the impact of different order categories over 30 minute intervals and Bouchaud et al. \cite{eisler10} who examine the impact of market orders, limit orders and cancelations at the level of individual events.

\subsection{Summary}
We conduct in this study an empirical investigation of the impact of order book events --market orders, limit orders and cancelations-- on equity prices.
Although previous studies give a relatively complex description of this impact, we argue that, in fact, their impact on  price dynamics may be modeled parsimoniously through a {\it single}
variable, the {\it order flow imbalance} (OFI),   which represents the net order flow at the bid and ask and tracks changes in the size of the bid and ask queues by
\begin{itemize}\item increasing every time the bid size increases, the ask size decreases or the bid/ask prices increase
\item decreases every time the bid size decreases, the ask size increases or the bid/ask prices decrease.\end{itemize}
    Interestingly, this variable treats a market sell and a cancel buy of the same size as equivalent, since they have the same effect on the size of the bid queue.
We find that this aggregate variable explains mid-price changes over short time scales in a linear fashion,  for a large sample of stocks, with an average $R^2$ of 65\%. The resulting price impact model relates prices, trades, limit orders and cancelations in a simple way: it is {\it linear}, requires the estimation of a single  parameter and it is robust across stocks and across timescales.

The slope of this relation, which we call the {\it price impact coefficient}, 
exhibits intraday seasonality in line with known intraday patterns observed in spreads, market depth and price volatility \cite{ahn01,andersen98,lee93,mcinish92} which have been explained in terms of intraday shifts in information asymmetry \cite{madhavan97} or informativeness of trades \cite{hasbrouck91-2}.
Motivated by a stylized model of the order book, we relate the intraday changes in the price impact coefficient to variations in market depth and show that price impact is inversely proportional to the depth of the order book.
This allows us to explain  intraday patterns in price impact and price volatility using only observable quantities - the order flow imbalance and the market depth, as opposed to  unobservable parameters previously invoked in the literature, such as information asymmetry or informativeness of trades.


The intuition that ``it takes volume to move prices", though widely confirmed by empirical studies \cite{karpoff87}, is not easy to explain theoretically (see  \cite[Chapter 6.2]{ohara98}). 
 In Section \ref{volume.sec}, we show that our price impact model, together with a scaling argument,
  leads to an apparent ``square root" relation between price changes and trade volume, similar to some findings in the empirical literature \cite{clark73, richardson86}.  However, we argue that this relation is not robust and is
  a statistical artifact due to the aggregation of data.

\subsection{Outline}
The article is structured as follows. In Section \ref{model.sec}, motivated by a stylized model of the order book, we specify a parsimonious model that links stock price changes, order flow imbalance and market depth. Section \ref{empirical.sec} describes the trades and quotes data and estimation results for our model. There, we also show how intraday patterns in depth and order flow imbalance generate intraday patterns in price impact and price volatility. In Section \ref{volume.sec} we discuss the role of trading volume as an explanatory variable and show that order flow imbalance is more  effective in explaining price moves than variables based on trades. We also derive a scaling relation between order flow imbalance and traded volume and show how the ``square-root'' price impact of volume follows from our model. We present our conclusions in Section \ref{conclusion.sec}.

\section{A model for the price impact of orders}\label{model.sec}

\subsection{Variables}
We focus on `Level I order book': the limit orders sitting at the best bid and ask. Every observation of the bid and the ask consists of the bid price $P^B$, the size $q^B$ of the bid queue (in number of shares), the ask price $P^A$ and the size $q^A$ of the ask queue (in number of shares):
\begin{figure}[ht]
\noindent
\begin{center}
{
\includegraphics[width=45mm]{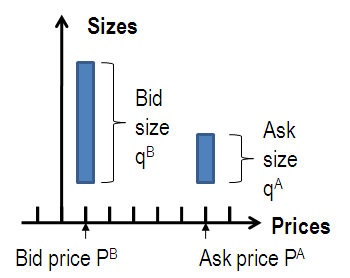}
}
\end{center}
\label{BBO}
\end{figure}

The bid price and size represent the demand for a stock, while the ask price and size represent the supply. We enumerate these observations by $n$ and compare $(P^B_{n-1},q^B_{n-1},P^A_{n-1},q^A_{n-1})$ with $(P^B_n,q^B_n,P^A_n,q^A_n)$. Between two such observations, only one of the following events can occur:
\begin{itemize}
\item $P^B_n>P^B_{n-1}$ or $q^B_n>q^B_{n-1}$ signifying an increase in demand
\item $P^B_n<P^B_{n-1}$ or $q^B_n<q^B_{n-1}$ signifying a decrease in demand
\item $P^A_n<P^A_{n-1}$ or $q^A_n>q^A_{n-1}$ signifying an increase in supply
\item $P^A_n>P^A_{n-1}$ or $q^A_n<q^A_{n-1}$ signifying a decrease in supply
\end{itemize}
We define the variable $e_n$ which  measures the contribution of the $n-$th event to the size of bid and ask queues:
$$e_n=I_{\{P^B_n\geq P^B_{n-1}\}}q^B_n-I_{\{P^B_n\leq P^B_{n-1}\}}q^B_{n-1}-I_{\{P^A_n\leq P^A_{n-1}\}}q^A_n+I_{\{P^A_n\geq P^A_{n-1}\}}q^A_{n-1}
$$
Note that if $q^B$ increases but $P^B$ remains the same, we assign $e_n=q^B_{n}-q^B_{n-1}$, representing the size that was added  at the bid. If $q^B$ decreases, we also assign $e_n=q^B_{n}-q^B_{n-1}$, representing the size that was removed from the bid, whether due to a market sell or cancel buy order.
If $P^B$ increases, we let $e_n=q^B_n$, representing the size of a price-improving limit order. If $P^B$ decreases, we let $e_n=q^B_{n-1}$, representing the size that was removed, whether due to a market order or a cancellation. The same classification is done for events on the ask side, with signs reversed.

Events affecting the order book occur at random times $\tau_n$, and we define $N(t) =\max\{n|\tau_n\leq t\}$ to be the number of events during $[0,t]$. We define the {\it order flow imbalance} over time intervals $[t_{k-1},t_{k}]$ as a sum of individual event contributions $e_n$ over these intervals:
$$OFI_{k}=\sum_{n=N(t_{k-1})+1}^{N(t_{k})}{e_n},$$
\noindent where $N(t_{k-1})+1$ and $N(t_{k})$ are the index of the first and the index of the last event in the interval $[t_{k-1},t_{k}]$. The order flow imbalance is a  measure of supply/demand imbalance, which encompasses trades, limit orders and cancelations. Whereas previous studies \cite{chordia08,hasbrouck91,hasbrouck01,kempf99,plerou02,torre97} focused on  measures of ``trade imbalance"\footnote{Hopman \cite{hopman07} computes the supply/demand imbalance based on limit orders and trades, but not cancelations.}, using orders provides a more natural way of measuring supply and demand.

We also consider mid-price changes (in number of ticks) over the same time grid:
$$\Delta P_{k}=(P_{k}-P_{k-1})/\delta,$$
\noindent where $P_{k}$ is the mid-quote price at time $t_k$ and $\delta$ is the tick size (equal to 1 cent in our data).

\subsection{A stylized  model of the order book}
Consider first a stylized model of the order book in which
\begin{enumerate}
\item the number of shares at each price level beyond the best bid/ask is equal to $D$.
\item limit orders arrivals and cancelations occur only at the best bid/ask.
\end{enumerate}
We will show that under these assumptions a linear relation holds between order flow imbalance and price changes.
Consider three scenarios, when only market buy orders, limit buy orders or limit sell cancels happen over some time interval $[t,t+\Delta t]$:
\begin{itemize}
\item Market sell orders remove $M^s$ shares from the bid.
\begin{figure}[h!]
\noindent
\begin{center}
{
\includegraphics[width=0.55\textwidth]{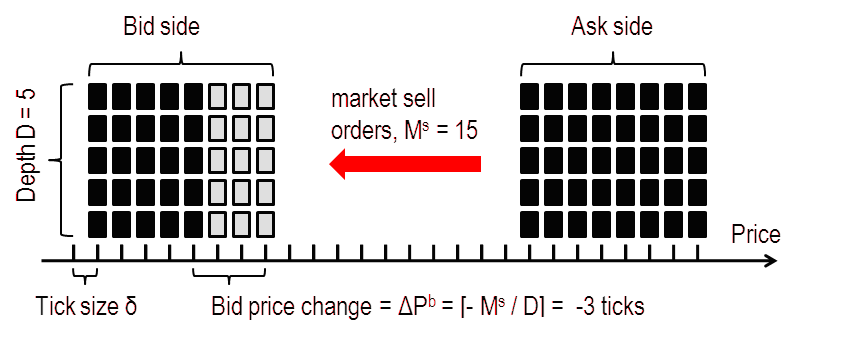}
}
\end{center}
\label{marketo}
\end{figure}
\item Market sell orders remove $M^s$ shares from the bid, while limit buy orders add $L^b$ shares to the bid.
\begin{figure}[h!]
\noindent
\begin{center}
{
\includegraphics[width=0.55\textwidth]{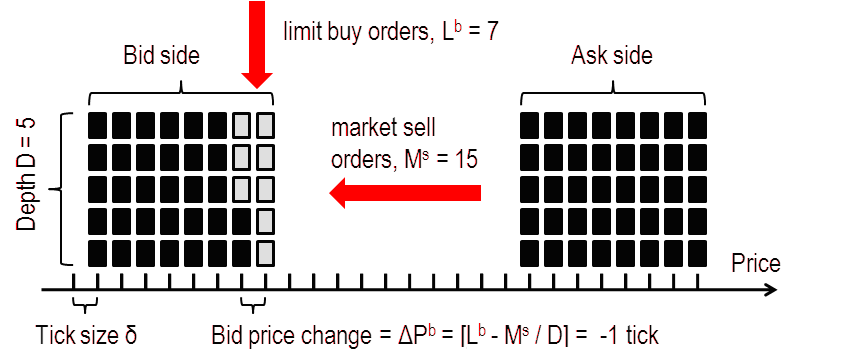}
}
\end{center}
\label{cancelo}
\end{figure}

\item Market sell orders and limit buy cancels remove $M^s+C^b$ shares from the bid, while limit buy orders add $L^b$ shares to the bid.
\begin{figure}[h!]
\noindent
\begin{center}
{
\includegraphics[width=0.55\textwidth]{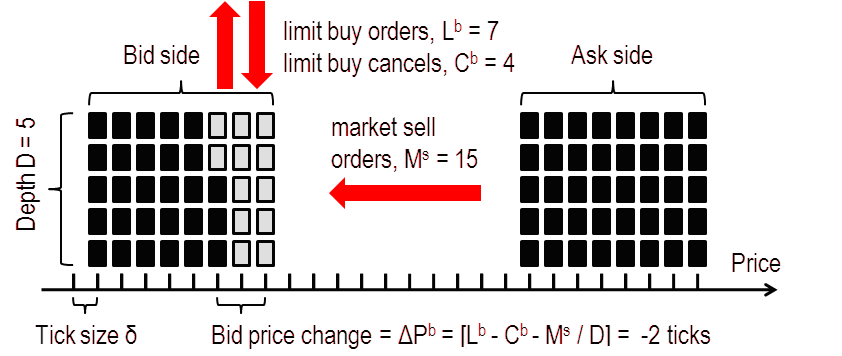}
}
\end{center}
\label{limito}
\end{figure}
\end{itemize}

The three variables $M^b$, $C^s$ and $L^s$ for the ask can be defined analogously. Under the above assumptions, the impact of order book events at the bid (ask) side of the book is {\it additive} and only depends on their net effect on the bid (ask) queue size:

$$\Delta P^b= \lceil (L^b-C^b-M^s)/D \rceil$$

Similarly, for the ask:

$$\Delta P^a= -\lceil (L^s-C^s-M^b)/D \rceil$$

These relations are remarkably simple - they involve no parameters and incorporate the effects of all order book events on bid and ask prices. Although the following analysis can be carried for the bid and the ask prices separately, we take their average (the mid-price) to simplify the analysis:

$$\Delta P= \frac{1}{2}\lceil (L^b-C^b-M^s)/D \rceil - \frac{1}{2}\lceil (L^s-C^s-M^b)/D \rceil$$

Note that the above is equivalent (up to truncation) to

\begin{equation}\label{toy3}
\Delta P=\frac{OFI}{2D}+\epsilon,
\end{equation}

where $OFI=L^b-C^b-M^s-L^s+C^s+M^b$ and $\epsilon$ is the truncation error. This expression for $OFI$ is obtained from its definition by grouping individual order contrubutions $e_i$ by their types (limit buys, market sells, etc). 

\subsection{Model specification}

In reality, order books have complex dynamics and the relation (\ref{toy3}) will only hold in a statistical sense. For example, limit orders and cancelations  occur at all levels of the order book. The distribution of depth across price levels often has humps, gaps and is itself a separate object of study  \cite{potters03,zovko02}. Moreover, the depth is subject to important intraday fluctuations. Finally, there may be hidden  orders in the book which  are not reported in the data \cite{avellaneda10}. With these considerations, we suggest the following relation:
\begin{equation}\label{simplespec}
\Delta P_{k}=\beta \quad OFI_{k}+\epsilon_{k},
\end{equation}
\noindent where $\beta$ is  the {\it price impact coefficient} and $\epsilon_{k}$ is a noise term due to the influence of deeper levels of the order book and rounding errors. Our earlier discussion suggests that the price impact coefficient is inversely related to market depth, which is itself subject to intraday fluctuations.
We define a measure of depth by averaging the bid/ask queue sizes over  intervals $[T_{i-1},T_{i}]$:
$$ AD_{i}=\frac{1}{2(N(T_{i})-N(T_{i-1})-1)}\sum_{n=N(T_{i-1})+1}^{N(T_{i})}{(q^B_n+q^A_n)}$$
We therefore specify the following relation between the price impact coefficient $\beta_{i}$  in the time interval $[T_{i-1},T_{i}]$  
and our measure of market depth as:
\begin{equation}\label{beta_depth_rel1}
\beta_i=\frac{c}{AD_{i}^\lambda} +\nu_{i},
\end{equation}
\noindent where $c,\lambda$ are constants 
and $\nu_{i}$ is a noise term. Note that the stylized model exposed above corresponds to $\lambda=1$.

The specification (\ref{simplespec}-\ref{beta_depth_rel1}) may be regarded as a model of the instantaneous price impact over a short time interval $[t_{k-1},t_k]$. An order, submitted at $\tau\in[t_{k-1},t_k]$, has a contribution $e_\tau$ and joins the aggregate order flow imbalance $OFI_{k}$. If the order goes in the same direction as the majority of the orders ($sgn(e_\tau)=sgn(OFI_{k})$), it reinforces the concurrent order flow imbalance and can affect the price. If the order goes against the concurrent order flow imbalance ($sgn(e_\tau)=-sgn(OFI_{k})$), it is compensated by other orders and may have an instantaneous impact of zero. In our model all events (including trades) have a linear price impact, equal to $\beta$ on average. Their realized impact, however, depends on the rest of the orders that arrive during the same time interval.

\begin{figure}[h!]
\noindent
\begin{center}
{
\includegraphics[width=0.5\textwidth]{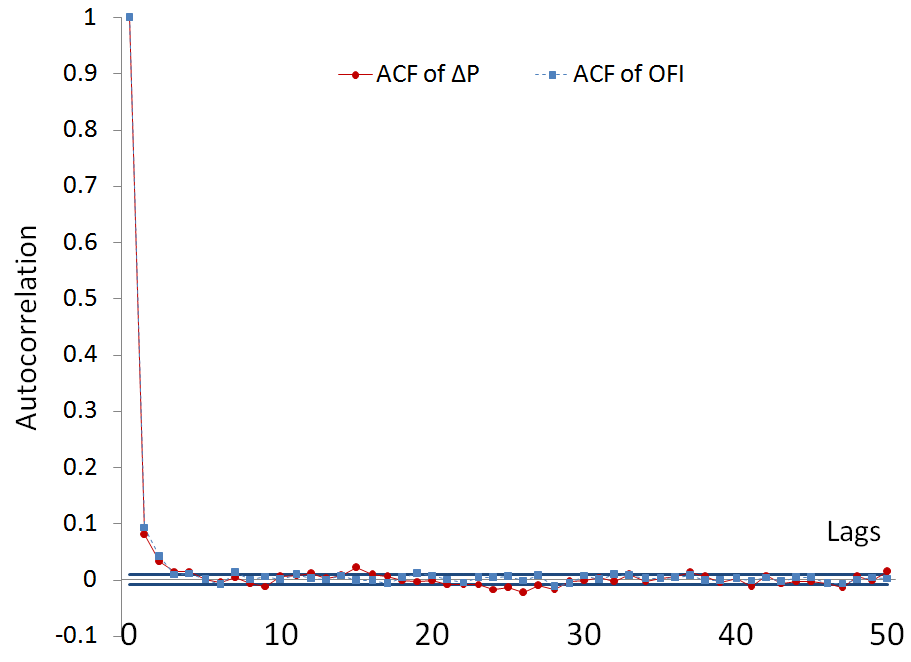}
}
\end{center}
\caption{ACF of the mid-price changes $\Delta P_{k}$, the order flow imbalance $OFI_{k}$ and the 5\% significance bounds for the Schlumberger stock (SLB).}
\label{acf0}
\end{figure}

The idea that the concurrent limit order activity can make a difference in terms of trades' impact was demonstrated by Stephens et al. \cite{stephens09}, where authors show that the shape of the price impact function essentially depends on the contemporaneous limit order activity. Our approach can also be related to the  model proposed by Bouchaud et al. \cite{eisler10}. where order book events have a linear impact on prices, which depends on their signs and types\footnote{Note that in our case all order book events have the same average impact, equal to $\beta_i$, regardless of their type. As shown in \cite{eisler10}, average impacts of different event types are empirically very similar, allowing to reasonably approximate them with a single number.}. The major difference of our models lies in the aggregation across time and events. As argued in \cite{eisler10}, order book events have complicated auto- and cross-correlation structures on the timescale of individual events, which typically vanish after 10 seconds. In our data the autocorrelations at a timescale of 10 seconds are small and quickly vanish as well (ACF plots for a representative stock are shown on Figure \ref{acf0}). Finally, Hasbrouck and Seppi \cite{hasbrouck01} propose a model similar to (\ref{simplespec}, \ref{beta_depth_rel1}) for explaining the price impact of trades. Although their focus is on trades, they also allow the price impact coefficient to depend on contemporaneous liquidity factors and change through time.

However, the linear equation (\ref{simplespec}) is quite different from models of price impact that consider only the size of trades \cite{gabaix03,hasbrouck91,kempf99,torre97,plerou02,potters03}. Instead of modeling price impact of trades as a (nonlinear) function of trade size, we show that the price impact of all events (including trades) is a linear function of their size after events are aggregated into a single imbalance variable. In Section \ref{volume.sec} we will argue that, first, the effect of trades on prices is adequately captured by the order flow imbalance and, second, that if one leaves out all events except trades, the relation \ref{simplespec} leads to an apparent concave relation between price changes and trade volume.

The next section provides an overview of the estimation results for our model.

\newpage

\section{Estimation and results}\label{empirical.sec}

\subsection{The trades and quotes (TAQ) data}

Our data set consists of one calendar month (April, 2010) of trades and quotes (TAQ) data for 50 stocks. The stocks were selected by a random number generator from the S\&P 500 constituents. The S\&P 500 composition for that month was obtained from Compustat and the data for individual stocks was obtained from the TAQ consolidated quotes and TAQ consolidated trades databases. The data were obtained through Wharton Research Data Services (WRDS).

Consolidated quotes contains all changes in queue sizes at the best bid and ask. For each stock, a data update consists of a timestamp (rounded to the nearest second), bid price, bid size, ask price, ask size and exchange flag. Consolidated trades (or market orders) consist of a timestamp, a price and a size. These two data sets are often referred to as Level 1 data, as opposed to Level 2 data, which also includes quote updates deeper in the book.

Our reason for using TAQ data rather than Level 2 order book data, is that it is far more accessible, yet contains all events in the top order book (best bid and ask updates). We demonstrate that Level 1 TAQ data can be successfully used to study limit orders and we hope that more empirical studies of that subject will follow. 
We note that the ratio of the number of quote updates to the number trades is roughly 40 to 1 in our data. Many empirical studies have focused exclusively on trades rather than quotes, but the sheer ratio in the size of these data sets is a good indicator that more information may be conveyed by the quotes than by trades.

Using a procedure described in detail in the appendix, we aggregate all quote updates to estimate the National Best Bid and Offer sizes and prices (NBBO) at each quote update.
Instead of aggregating all exchanges in this fashion, one may also simply filter by the exchange flag and study one exchange at the time. Focussing on one exchange at a time yields similar results.

We use a uniform grid in time $\{t_0,\dots,t_N\}$ with a timescale $t_{k}-t_{k-1}\equiv\Delta t=10$ seconds to compute the price changes and the order flow imbalances. To test the robustness of our findings to the choice of the basic timescale, we repeated our calculations on a subsample of stocks for different values of $\Delta t$, ranging from 10 quote updates (usually less than half of a second in our data) up to 10 minutes. The fit of our model generally increases with $\Delta t$, but the rest of the results stays the same. Time aggregation serves two purposes: first, it alleviates the issue of data discreteness and second, it mitigates the errors due to the trade matching algorithm (described in the Appendix).

\subsection{Empirical findings}

\begin{figure}[h!]
\noindent
\begin{center}
{
\includegraphics[width=100mm]{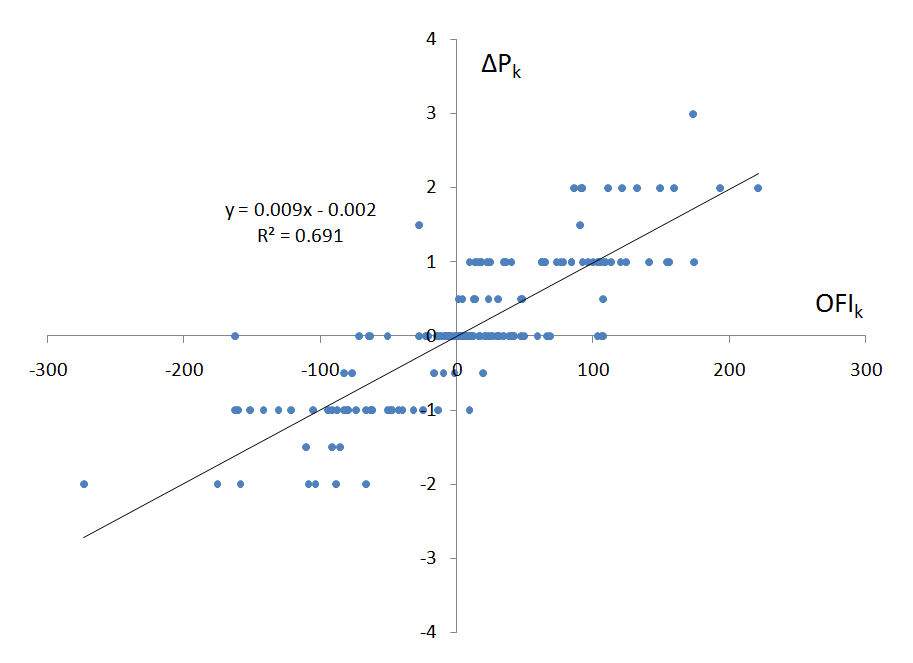}
}
\end{center}
\caption{Scatter plot of $\Delta P_{k}$ against $OFI_{k}$ for the Schlumberger stock (SLB), 04/01/2010 11:30-12:00pm.}
\label{regr_SLB}
\end{figure}

\begin{figure}[h!]
\noindent
\begin{center}
{
\includegraphics[width=100mm]{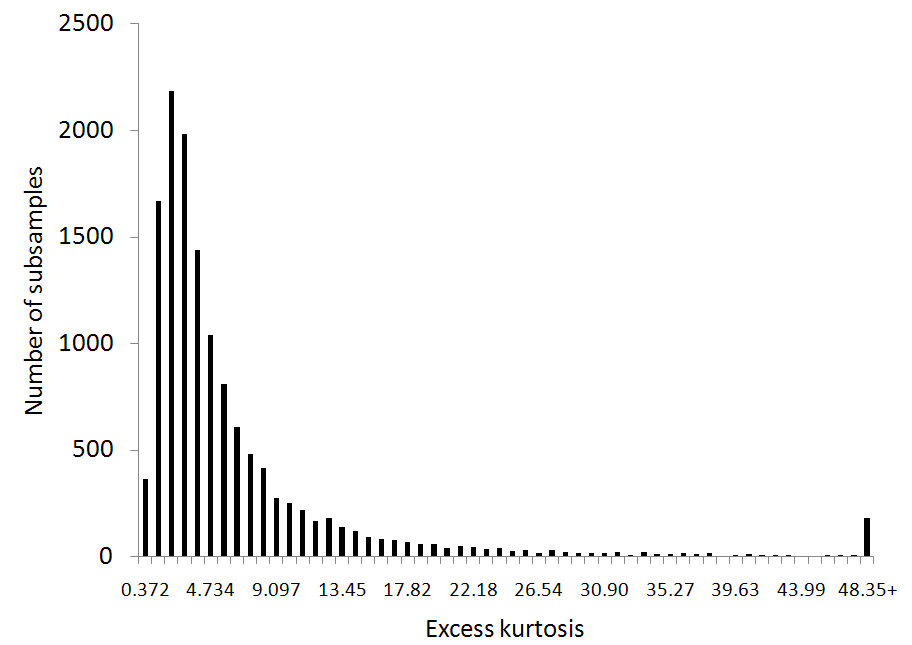}
}
\end{center}
\caption{Distribution of excess kurtosis of the residuals $\hat{\epsilon}_{k}$ across stocks and time.}
\label{kurt}
\end{figure}

We assume that the price impact coefficient $\beta$ is  constant over  each half-hour interval $[T_i,T_{i+1}]$ and estimate the model by ordinary least squares  regression  in each half-hour subsample for each stock:
\begin{equation}\label{simpleregr}
\Delta P_{k}=\hat{\alpha}_{i}+\hat{\beta}_{i}OFI_{k}+\hat{\epsilon}_{k},
\end{equation}
Figure \ref{regr_SLB} presents a scatter plot of $\Delta P_{k}$ against $OFI_{k}$ for one of the half-hour subsamples.
 Table 2 reports regression outputs, averaged across time for each stock. This table provides strong evidence of a linear relation between order flow imbalance and price changes. The goodness of fit is surprisingly high for all of the stocks, suggesting that the model (\ref{simplespec}) performs well regardless of stock-specific features\footnote{ We note that $OFI_{k}$ includes the contributions of price-changing order book events, leading to a possible tautology in the regression (\ref{simpleregr}). This problem is inherent to all price impact modeling, because the explanatory variables (events or trades) can directly cause price changes. To test that the high $R^2$ in our regressions is not due to this tautology, we estimated (\ref{simpleregr}) on a subsample of stocks, excluding the price-changing events from $OFI_{k}$. With this change the $R^2$ declined, but remained in the 35\%-60\% region.}. In addition to the high quality of fits, the regression coefficient $\beta_i$ is virtually always statistically significant (at a 95\% level of the z-test), while the intercept is mostly insignificant. Figure \ref{kurt} represents the histogram of excess kurtosis values of the residuals $\hat{\epsilon}_{k}$ across subsamples: the relatively low level of kurtosis shows that the residuals are not predominantly associated with large price changes.
  Since the regression residuals demonstrate heteroscedasticity, we used White's heteroscedasticity-consistent standard errors for the z-test. To check for higher order/nonlinear dependence we add a quadratic term $\hat{\gamma}_{Q,i}OFI_{k}|OFI_{k}|$ to the regression. The  increase in $R^2$, from 65\% to 68\% on average, is
 barely noticeable and the coefficient $\hat{\gamma}_{Q,i}$ is statistically insignificant in most samples.

\newpage

\begin{center}
{Table 1. Descriptive statistics}
\vskip 6pt
\scriptsize
\begin{tabular}{|l|l|r|r|r|r|r|r|r|}
\hline
\multirow{3}{*}{Name} & \multirow{3}{*}{Ticker} & \multirow{3}{*}{Price} & Daily & Number of & Number of & Average & Maximum & Best quote\\
 &  &  & volume, & best quote & trades & Spread, & spread, & depth,\\
 &  &  & shares & updates &  & cents & cents & shares\\
\hline
Advanced Micro Devices & AMD  & 9.61 & 20872996 & 417204 & 6687 & 1 & 1 & 1035\\

Apollo Group & APOL & 62.92 & 1949337 & 172942 & 4095 & 2 & 5 & 15\\

American Express & AXP  & 45.21 & 8678723 & 559701 & 7748 & 1 & 24 & 79\\

Autozone & AZO  & 179.03 & 243197 & 43682 & 1081 & 9 & 35 & 7\\

Bank of America & BAC  & 18.43 & 164550168 & 1529395 & 15008 & 1 & 1 & 3208\\

Becton Dickinson & BDX  & 78.07 & 1130362 & 61029 & 2968 & 2 & 5 & 15\\

Bank of New York Mellon & BK   & 31.77 & 6310701 & 285619 & 5518 & 1 & 1 & 122\\

Boston Scientific & BSX  & 7.13 & 25746787 & 309441 & 6768 & 1 & 1 & 2965\\

Peabody Energy corp & BTU  & 47.14 & 5210642 & 298616 & 7267 & 1 & 3 & 29\\

Caterpillar & CAT  & 67.20 & 6664891 & 392499 & 8224 & 1 & 2 & 38\\

Chubb & CB   & 52.22 & 1951618 & 149010 & 3601 & 1 & 2 & 43\\

Carnival & CCL  & 40.16 & 4275911 & 215427 & 5503 & 1 & 2 & 53\\

Cincinnati Financial & CINF & 29.41 & 688914 & 51373 & 1528 & 1 & 2 & 42\\

CME Group & CME  & 322.83 & 418955 & 38504 & 1412 & 31 & 103 & 5\\

Coach & COH  & 41.91 & 3126469 & 176795 & 4458 & 1 & 2 & 41\\

ConocoPhillips & COP  & 56.09 & 9644544 & 426614 & 8621 & 1 & 2 & 84\\

Coventry Health Care & CVH  & 24.16 & 1157022 & 79305 & 2213 & 1 & 2 & 38\\

Denbury Resources & DNR  & 17.88 & 5737740 & 263173 & 4643 & 1 & 1 & 186\\

Devon Energy & DVN  & 66.98 & 3260982 & 177006 & 5805 & 2 & 4 & 18\\

Equifax & EFX  & 35.34 & 799505 & 62957 & 1945 & 1 & 3 & 39\\

Eaton & ETN  & 78.53 & 1757136 & 67989 & 3580 & 2 & 6 & 13\\

Fiserv & FISV & 52.56 & 1038311 & 58304 & 2208 & 1 & 3 & 20\\

Hasbro & HAS  & 39.48 & 1322037 & 86040 & 2672 & 1 & 2 & 34\\

HCP & HCP  & 32.63 & 2872521 & 213045 & 4357 & 1 & 2 & 48\\

Starwood Hotels & HOT  & 50.59 & 3164807 & 150252 & 5106 & 2 & 4 & 22\\

Kohl's & KSS  & 56.88 & 3064821 & 128196 & 4936 & 1 & 3 & 27\\

L-3 Communications & LLL  & 94.64 & 670937 & 72818 & 2141 & 2 & 6 & 9\\

Lockheed Martin & LMT  & 84.14 & 1416072 & 88254 & 3333 & 2 & 5 & 15\\

Macy's & M    & 23.40 & 8324639 & 491756 & 6469 & 1 & 1 & 176\\

Marriott & MAR  & 34.45 & 5014098 & 238190 & 5499 & 1 & 2 & 65\\

McAfee & MFE  & 40.04 & 2469324 & 109073 & 3561 & 1 & 2 & 40\\

McGraw-Hill & MHP  & 34.90 & 1954576 & 102389 & 3261 & 1 & 2 & 42\\

Medco Health Solutions & MHS  & 63.22 & 2798098 & 109382 & 4680 & 1 & 3 & 25\\

Merck & MRK  & 36.03 & 13930842 & 448748 & 7997 & 1 & 1 & 231\\

Marathon Oil & MRO  & 32.33 & 5035354 & 341408 & 5522 & 1 & 1 & 143\\

MeadWestvaco & MWV  & 26.96 & 1035547 & 92825 & 2312 & 1 & 3 & 37\\

Newmont Mining & NEM  & 53.43 & 5673718 & 435295 & 7717 & 1 & 2 & 38\\

Omnicom & OMC  & 41.17 & 3357585 & 150800 & 4359 & 1 & 2 & 65\\

MetroPCS Communications & PCS  & 7.53 & 4424560 & 107967 & 2901 & 1 & 1 & 523\\

Pultegroup & PHM  & 11.80 & 6834683 & 262420 & 4604 & 1 & 1 & 319\\

PerkinElmer & PKI  & 23.98 & 1268774 & 78114 & 2127 & 1 & 2 & 72\\

Ryder System & R    & 44.01 & 631889 & 47422 & 2085 & 2 & 5 & 11\\

Reynolds American & RAI  & 54.44 & 773387 & 56236 & 2076 & 1 & 4 & 22\\

Schlumberger & SLB  & 67.94 & 9476060 & 440839 & 10286 & 1 & 2 & 39\\

Teco Energy & TE   & 16.52 & 1070815 & 70318 & 1807 & 1 & 1 & 148\\

Time Warner Cable & TWC  & 53.21 & 1770234 & 88286 & 3554 & 2 & 3 & 22\\

Whirlpool & WHR  & 97.73 & 1424264 & 134152 & 3348 & 4 & 9 & 10\\

Windstream & WIN  & 11.03 & 2508830 & 104887 & 2937 & 1 & 1 & 798\\

Watson Pharmaceuticals & WPI  & 42.51 & 895967 & 63094 & 2024 & 1 & 3 & 29\\

XTO Energy & XTO & 48.13 & 7219436 & 612804 & 5040 & 1 & 7 & 225\\
\hline
Grand mean &  & 51.75 & 7512376 & 223232 & 4552 & 2 & 6 & 227\\
\hline
\end{tabular}
\end{center}
\vskip 12pt

Table 1 presents the average mid-price, daily transaction volume, daily number of best quote updates, daily number of trades, spread and the depth at the best bid and ask for 50 randomly chosen U.S. stocks. All values are calculated from the filtered data, that consists of 21 trading day during April, 2010.
\normalsize

\newpage

\begin{center}
{Table 2. Relation between price changes and order flow imbalance.}
\scriptsize
\vskip 6pt
\begin{tabular}{|l|r|r|r|r|r|r|r|r|r|r|}
\hline
\multirow{2}{*}{Ticker} & \multicolumn{7}{|c|}{Average results} & \multicolumn{3}{|c|}{Hypothesis testing}\\
\cline{2-11}
 & $\hat{\alpha}$ & $t(\hat{\alpha})$ & $\hat{\beta}$ & $t(\hat{\beta})$ & $\hat{\gamma}_Q$ & $t(\hat{\gamma}_Q)$ & $R^2$ & $\{\alpha\neq 0\}$ & $\{\beta\neq 0\}$ & $\{\gamma_Q\neq 0\}$\\
\hline

AMD  & -0.0032 & -0.17 & 0.0008 & 9.96 & 1.4E-07 & 0.68 & 64\% & 0\% & 98\% & 22\%\\

APOL & 0.0038 & 0.10 & 0.0555 & 10.32 & -2.2E-04 & -1.17 & 63\% & 12\% & 91\% & 4\%\\

AXP  & 0.0019 & 0.08 & 0.0082 & 13.87 & -3.8E-06 & -0.88 & 69\% & 11\% & 100\% & 5\%\\

AZO  & 0.0101 & 0.33 & 0.1619 & 6.39 & -9.3E-04 & -0.89 & 47\% & 23\% & 97\% & 3\%\\

BAC  & -0.0018 & -0.09 & 0.0002 & 18.36 & 1.9E-09 & 0.01 & 79\% & 1\% & 100\% & 8\%\\

BDX  & -0.0008 & -0.06 & 0.0536 & 10.08 & -1.1E-04 & -0.38 & 63\% & 9\% & 100\% & 8\%\\

BK   & -0.0078 & -0.19 & 0.0069 & 14.97 & -4.0E-06 & -0.57 & 74\% & 3\% & 100\% & 6\%\\

BSX  & 0.0000 & -0.01 & 0.0003 & 6.12 & 7.8E-08 & 1.14 & 58\% & 0\% & 81\% & 22\%\\

BTU  & 0.0048 & 0.12 & 0.0242 & 14.51 & -3.5E-05 & -1.26 & 72\% & 11\% & 100\% & 3\%\\

CAT  & 0.0147 & 0.23 & 0.0194 & 14.85 & -1.9E-05 & -1.13 & 71\% & 12\% & 99\% & 3\%\\

CB   & -0.0086 & -0.07 & 0.0191 & 11.97 & -3.5E-07 & 0.00 & 64\% & 5\% & 100\% & 8\%\\

CCL  & -0.0067 & -0.18 & 0.0140 & 13.88 & -1.2E-05 & -0.64 & 70\% & 3\% & 99\% & 7\%\\

CINF & -0.0030 & -0.02 & 0.0260 & 10.73 & -7.0E-06 & 0.27 & 70\% & 1\% & 98\% & 16\%\\

CME  & 0.0506 & 0.05 & 0.6262 & 4.98 & -7.2E-03 & -0.99 & 35\% & 15\% & 94\% & 2\%\\

COH  & -0.0221 & -0.45 & 0.0179 & 12.75 & -1.7E-05 & -0.77 & 69\% & 2\% & 100\% & 3\%\\

COP  & -0.0008 & 0.06 & 0.0084 & 12.50 & -5.8E-06 & -1.17 & 68\% & 10\% & 100\% & 3\%\\

CVH  & -0.0034 & -0.06 & 0.0217 & 10.83 & 7.6E-06 & 0.20 & 65\% & 3\% & 99\% & 10\%\\

DNR  & -0.0008 & -0.04 & 0.0045 & 12.76 & -1.3E-07 & 0.19 & 69\% & 1\% & 99\% & 13\%\\

DVN  & 0.0112 & 0.18 & 0.0370 & 11.48 & -1.0E-04 & -1.59 & 65\% & 17\% & 97\% & 0\%\\

EFX  & -0.0032 & -0.04 & 0.0222 & 8.71 & 6.4E-05 & 0.64 & 56\% & 1\% & 98\% & 18\%\\

ETN  & -0.0076 & 0.05 & 0.0712 & 10.51 & -2.3E-04 & -1.14 & 65\% & 14\% & 98\% & 1\%\\

FISV & -0.0002 & 0.06 & 0.0397 & 10.42 & -2.3E-05 & -0.19 & 63\% & 4\% & 100\% & 8\%\\

HAS  & -0.0031 & -0.02 & 0.0222 & 11.45 & 4.7E-06 & 0.21 & 67\% & 3\% & 100\% & 16\%\\

HCP  & -0.0078 & -0.17 & 0.0150 & 13.60 & -1.4E-05 & -0.46 & 67\% & 2\% & 100\% & 6\%\\

HOT  & -0.0012 & 0.05 & 0.0345 & 12.64 & -7.2E-05 & -1.21 & 68\% & 10\% & 99\% & 2\%\\

KSS  & -0.0030 & -0.04 & 0.0317 & 13.82 & -5.4E-05 & -0.80 & 71\% & 10\% & 98\% & 3\%\\

LLL  & 0.0160 & 0.32 & 0.1000 & 11.76 & -3.8E-04 & -0.75 & 67\% & 14\% & 96\% & 3\%\\

LMT  & 0.0006 & 0.00 & 0.0520 & 13.58 & -1.2E-04 & -0.98 & 72\% & 14\% & 100\% & 1\%\\

M    & -0.0010 & 0.04 & 0.0043 & 15.82 & 8.8E-08 & 0.13 & 75\% & 0\% & 100\% & 12\%\\

MAR  & -0.0039 & -0.02 & 0.0121 & 14.61 & -4.1E-06 & -0.23 & 71\% & 3\% & 100\% & 4\%\\

MFE  & 0.0087 & 0.16 & 0.0205 & 12.72 & -3.8E-05 & -0.38 & 68\% & 7\% & 100\% & 7\%\\

MHP  & -0.0073 & -0.13 & 0.0211 & 11.62 & 5.8E-06 & 0.14 & 68\% & 2\% & 99\% & 11\%\\

MHS  & -0.0055 & -0.16 & 0.0334 & 11.70 & -8.3E-05 & -1.10 & 66\% & 9\% & 99\% & 3\%\\

MRK  & -0.0065 & -0.20 & 0.0032 & 12.53 & -5.4E-07 & -0.38 & 69\% & 1\% & 100\% & 8\%\\

MRO  & 0.0018 & 0.07 & 0.0058 & 13.67 & -3.6E-07 & 0.22 & 69\% & 5\% & 100\% & 13\%\\

MWV  & -0.0011 & 0.01 & 0.0205 & 11.79 & -1.7E-05 & -0.25 & 68\% & 3\% & 100\% & 7\%\\

NEM  & -0.0102 & -0.22 & 0.0170 & 13.81 & -1.9E-05 & -1.36 & 71\% & 8\% & 100\% & 2\%\\

OMC  & -0.0099 & -0.28 & 0.0144 & 11.88 & -4.5E-06 & -0.01 & 65\% & 2\% & 99\% & 13\%\\

PCS  & -0.0006 & -0.03 & 0.0015 & 5.21 & 1.8E-06 & 1.01 & 53\% & 0\% & 79\% & 24\%\\

PHM  & 0.0006 & 0.03 & 0.0027 & 10.33 & 8.4E-07 & 0.55 & 66\% & 1\% & 98\% & 21\%\\

PKI  & -0.0004 & -0.03 & 0.0102 & 7.25 & 4.1E-05 & 1.10 & 53\% & 2\% & 94\% & 29\%\\

R    & 0.0006 & 0.03 & 0.0667 & 10.14 & 3.7E-05 & 0.01 & 63\% & 8\% & 98\% & 10\%\\

RAI  & -0.0070 & -0.10 & 0.0396 & 10.40 & 2.6E-05 & 0.01 & 66\% & 5\% & 100\% & 11\%\\

SLB  & -0.0077 & -0.15 & 0.0198 & 16.76 & -1.8E-05 & -1.15 & 76\% & 7\% & 100\% & 1\%\\

TE   & 0.0011 & 0.05 & 0.0049 & 6.66 & 1.4E-05 & 1.45 & 54\% & 2\% & 86\% & 30\%\\

TWC  & -0.0130 & -0.13 & 0.0384 & 11.80 & -5.6E-05 & -0.44 & 64\% & 8\% & 99\% & 5\%\\

WHR  & 0.0628 & 0.63 & 0.1278 & 10.26 & -3.3E-04 & -0.80 & 65\% & 22\% & 97\% & 4\%\\

WIN  & -0.0004 & -0.03 & 0.0009 & 3.12 & 1.5E-06 & 0.76 & 44\% & 1\% & 60\% & 15\%\\

WPI  & -0.0090 & -0.21 & 0.0270 & 10.47 & 2.9E-05 & 0.28 & 66\% & 3\% & 98\% & 14\%\\

XTO  & -0.0088 & -0.18 & 0.0029 & 13.28 & 2.7E-07 & 0.30 & 65\% & 0\% & 100\% & 18\%\\
\hline
Average & 0.0002 & -0.02 & 0.0398 & 11.47 & -2.0E-04 & -0.28 & 65\% & 6\% & 97\% & 9\%\\
\hline
\end{tabular}
\end{center}
\vskip 12pt
\scriptsize

\noindent
Table 2 presents a cross-section of results (averaged across time) for the regressions:
\vskip 6pt
$\Delta P_{k}=\hat{\alpha}_{i}+\hat{\beta}_{i}OFI_{k}+\hat{\epsilon}_{k}$,

$\Delta P_{k}=\hat{\alpha}_{Q,i}+\hat{\beta}_{Q,i}OFI_{k}+ \hat{\gamma}_{Q,i}OFI_{k}|OFI_{k}|+\hat{\epsilon}_{Q,k}$,
\vskip 6pt

\noindent where $\Delta P_{k}$ are the 10-second mid-price changes and $OFI_{k}$ are the contemporaneous order flow imbalances. These regressions were estimated using 273 half-hour subsamples (indexed by $i$) for each stock and their outputs were averaged across subsamples. Each subsample typically contains about 180 observations (indexed by $k$). The t-statistics were computed using White's standard errors. For brevity, we report the $R^2$, the average $\hat{\alpha}_i$ and the average $\hat{\beta}_i$ only for the first regression (with a single $OFI_{k}$ term). There is almost no difference between averages of estimates $\hat{\beta}_i$ and $\hat{\beta_Q}_i$ and the $R^2$ in two regressions. The last three columns report the percentage of samples where the coefficient(s) passed the z-test at the 5\% significance level.
\normalsize

\newpage

Next, we estimate the parameters $\lambda$ and $c$ in (\ref{beta_depth_rel1}). For each stock, we first obtain $\hat{\lambda}$ fit via a loglinear regression:

\begin{equation}\label{beta_depth_regr}
\log{\hat{\beta}_i}=\hat{\alpha_{L,i}}-\hat{\lambda}\log{AD_i}+\hat{\epsilon}_{L,i}
\end{equation}

Then, using $\hat{\lambda}$, we estimate $c$ in a linear regression:

\begin{equation}\label{beta_c_regr}
\hat{\beta}_i=\hat{\alpha_{M,i}}+\frac{\hat{c}}{AD^{\hat{\lambda}}_i}+\hat{\epsilon}_{M,i}
\end{equation}

\begin{figure}[h!]
\noindent
\begin{center}
{
\includegraphics[width=120mm]{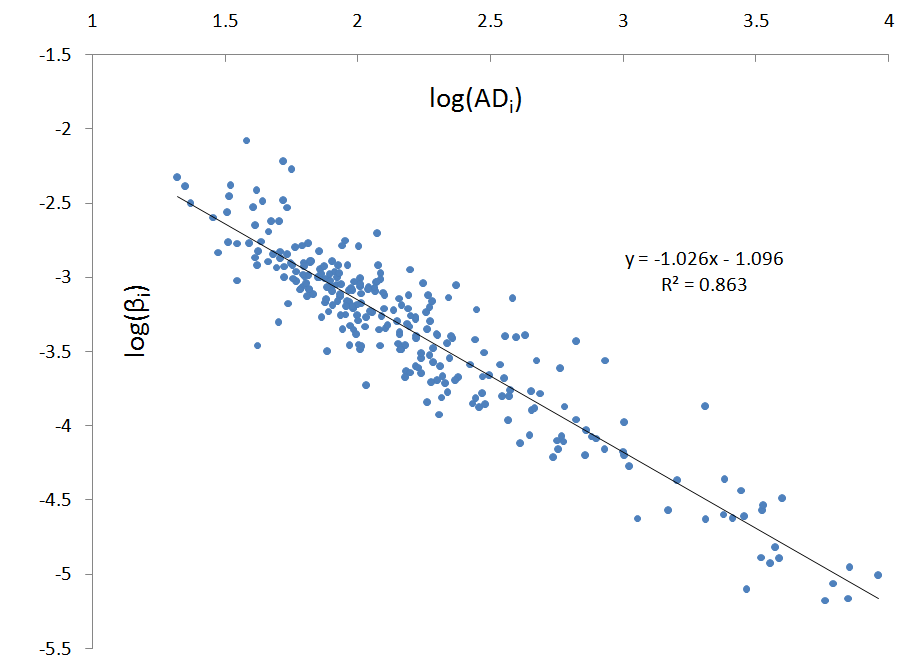}
}
\end{center}
\caption{Log-log scatter plot of the price impact coefficient estimate $\hat{\beta_i}$ against average market depth $AD_i$ for the Schlumberger stock (SLB).}
\label{betadepthSLB}
\end{figure}

Both regressions are estimated using ordinary least squares. The results are presented in Table 3: the quality of these fits convincingly demonstrates that the instantaneous price impact (measured by $\hat{\beta}_i$) is inversely related to market depth. There are three stocks with bad fits (namely APOL, AZO and CME) and we note that they also have wide spreads and low values of depth. It is possible that for these stocks other factors, such as the presence of hidden orders and depth beyond the best price levels the order book may dominate the instantaneous price impact. The intercept $\hat{\alpha_{L,i}}$ is highly statistically significant (being an estimate of parameter $c$) and $\hat{\alpha_{M,i}}$, which is included to absorb the means, is mostly insignificant. Since the residuals of these regressions appear to be autocorrelated, the t-statistics and confidence intervals in Table 3 are computed with Newey-West standard errors. Coinciding with our intuition for (\ref{toy3}), estimates $\hat{\lambda}$ are very close to 1 across stocks and the hypothesis $\{\lambda=1\}$ cannot be rejected for 35 out of 50 stocks. The restricted model (with $\lambda=1$) also demonstrates a good quality of fit, making this a good approximation. However, the coefficient $\hat{c}$ is generally different from $c=\frac{1}{2}$ in (\ref{toy3}). Lower values of $\hat{c}$ mean that mid-prices are (on average) more resilient to the incoming orders than indicated by $AD_i$ (which is only a rough measure of market depth). In summary, $\lambda=1$ appears to be a good approximation for most of the stocks and only the constant $c$ needs to be calibrated to the data. The general case of regression (\ref{beta_depth_regr}) is illustrated on Figure \ref{betadepthSLB} by a scatter plot for a representative stock.

\newpage

\begin{center}
{Table 3. Relation between the price impact coefficient and market depth.}
\scriptsize
\vskip 6pt
\begin{tabular}{|l|r|r|r|r|r|r|r|r|r|r|r|}
\hline
\multirow{2}{*}{Ticker} & \multicolumn{4}{|c|}{Parameter estimates} & \multicolumn{4}{|c|}{5\% confidence intervals} & \multicolumn{3}{|c|}{Fit measures}\\
\cline{2-12}
 & $\hat{c}$ & $\hat{\lambda}$ & $t({\hat{c}})$ & $t(\hat{\lambda})$ & $\hat{c}_l$ & $\hat{c}_u$ & $\hat{\lambda}_l$ & $\hat{\lambda}_u$ & $R^2$ & $corr[\hat{\beta},\hat{\hat{\beta}}]^2$ & $corr[\hat{\beta},\hat{\hat{\beta}}^*]^2$\\
\hline
AMD  & 0.23 & 0.94 & 27.74 & 23.11 & 0.22 & 0.25 & 0.86 & 1.02 & 78\% & 86\% & 86\%\\

APOL & 0.27 & 0.36 & 4.43 & 1.05 & 0.15 & 0.39 & -0.32 & 1.04 & 2\% & 30\% & 31\%\\

AXP  & 0.14 & 0.83 & 13.95 & 26.48 & 0.12 & 0.16 & 0.77 & 0.89 & 84\% & 76\% & 76\%\\

AZO  & 0.39 & 0.67 & 5.48 & 5.10 & 0.25 & 0.53 & 0.41 & 0.92 & 13\% & 17\% & 16\%\\

BAC  & 0.27 & 0.96 & 25.27 & 19.74 & 0.25 & 0.29 & 0.90 & 1.03 & 76\% & 87\% & 87\%\\

BDX  & 0.38 & 1.04 & 22.83 & 18.64 & 0.35 & 0.41 & 0.93 & 1.15 & 71\% & 68\% & 68\%\\

BK   & 0.21 & 0.92 & 17.52 & 54.54 & 0.19 & 0.24 & 0.88 & 0.95 & 93\% & 91\% & 90\%\\

BSX  & 0.35 & 0.98 & 14.98 & 24.55 & 0.31 & 0.40 & 0.90 & 1.05 & 73\% & 81\% & 81\%\\

BTU  & 0.42 & 1.12 & 40.90 & 36.77 & 0.40 & 0.44 & 1.06 & 1.18 & 87\% & 83\% & 83\%\\

CAT  & 0.29 & 0.96 & 21.70 & 16.87 & 0.27 & 0.32 & 0.85 & 1.07 & 87\% & 83\% & 83\%\\

CB   & 0.32 & 1.02 & 27.08 & 49.61 & 0.30 & 0.34 & 0.98 & 1.06 & 92\% & 89\% & 89\%\\

CCL  & 0.26 & 0.96 & 24.36 & 37.55 & 0.24 & 0.29 & 0.91 & 1.01 & 87\% & 83\% & 83\%\\

CINF & 0.31 & 0.97 & 20.05 & 47.39 & 0.28 & 0.34 & 0.93 & 1.01 & 92\% & 88\% & 88\%\\

CME  & 1.27 & 0.50 & 2.55 & 1.99 & 0.29 & 2.24 & 0.01 & 0.99 & 2\% & 4\% & 3\%\\

COH  & 0.37 & 1.05 & 15.29 & 36.65 & 0.32 & 0.43 & 0.98 & 1.12 & 77\% & 75\% & 75\%\\

COP  & 0.13 & 0.80 & 8.52 & 15.95 & 0.10 & 0.16 & 0.70 & 0.89 & 75\% & 66\% & 66\%\\

CVH  & 0.32 & 1.03 & 26.50 & 37.51 & 0.29 & 0.34 & 0.98 & 1.08 & 89\% & 89\% & 89\%\\

DNR  & 0.23 & 0.96 & 32.44 & 40.90 & 0.22 & 0.24 & 0.92 & 1.01 & 91\% & 89\% & 89\%\\

DVN  & 0.26 & 0.91 & 13.50 & 16.66 & 0.22 & 0.30 & 0.80 & 1.02 & 45\% & 56\% & 56\%\\

EFX  & 0.30 & 0.99 & 20.16 & 26.13 & 0.27 & 0.33 & 0.92 & 1.07 & 84\% & 79\% & 79\%\\

ETN  & 0.45 & 1.07 & 11.51 & 17.34 & 0.38 & 0.53 & 0.95 & 1.19 & 60\% & 56\% & 56\%\\

FISV & 0.34 & 1.01 & 23.35 & 30.70 & 0.31 & 0.36 & 0.94 & 1.07 & 84\% & 77\% & 77\%\\

HAS  & 0.32 & 1.00 & 26.36 & 46.00 & 0.30 & 0.34 & 0.96 & 1.05 & 89\% & 83\% & 83\%\\

HCP  & 0.19 & 0.89 & 22.93 & 51.27 & 0.17 & 0.21 & 0.86 & 0.93 & 94\% & 90\% & 90\%\\

HOT  & 0.44 & 1.12 & 19.53 & 26.59 & 0.40 & 0.48 & 1.04 & 1.20 & 82\% & 80\% & 79\%\\

KSS  & 0.39 & 1.05 & 24.40 & 33.17 & 0.36 & 0.42 & 0.99 & 1.11 & 85\% & 78\% & 78\%\\

LLL  & 0.43 & 1.01 & 13.21 & 14.45 & 0.37 & 0.50 & 0.87 & 1.14 & 51\% & 58\% & 58\%\\

LMT  & 0.50 & 1.14 & 7.31 & 13.49 & 0.37 & 0.64 & 0.98 & 1.31 & 60\% & 52\% & 52\%\\

M    & 0.19 & 0.90 & 37.41 & 57.39 & 0.18 & 0.20 & 0.87 & 0.93 & 94\% & 92\% & 92\%\\

MAR  & 0.28 & 0.98 & 22.58 & 50.20 & 0.25 & 0.30 & 0.94 & 1.02 & 92\% & 88\% & 88\%\\

MFE  & 0.31 & 1.01 & 20.28 & 46.20 & 0.28 & 0.34 & 0.96 & 1.05 & 91\% & 86\% & 86\%\\

MHP  & 0.27 & 0.94 & 19.60 & 33.62 & 0.24 & 0.30 & 0.89 & 1.00 & 82\% & 74\% & 74\%\\

MHS  & 0.53 & 1.16 & 17.03 & 34.25 & 0.47 & 0.59 & 1.10 & 1.23 & 85\% & 81\% & 80\%\\

MRK  & 0.13 & 0.81 & 18.07 & 32.20 & 0.11 & 0.14 & 0.76 & 0.86 & 87\% & 81\% & 81\%\\

MRO  & 0.23 & 0.94 & 35.54 & 49.68 & 0.21 & 0.24 & 0.91 & 0.98 & 94\% & 93\% & 93\%\\

MWV  & 0.32 & 1.05 & 28.07 & 37.81 & 0.30 & 0.34 & 1.00 & 1.10 & 90\% & 85\% & 85\%\\

NEM  & 0.26 & 0.98 & 18.79 & 25.97 & 0.23 & 0.28 & 0.91 & 1.05 & 81\% & 77\% & 77\%\\

OMC  & 0.30 & 0.96 & 29.47 & 17.76 & 0.28 & 0.32 & 0.85 & 1.06 & 83\% & 85\% & 85\%\\

PCS  & 0.30 & 1.02 & 21.27 & 18.73 & 0.27 & 0.33 & 0.90 & 1.14 & 53\% & 82\% & 82\%\\

PHM  & 0.28 & 0.98 & 36.43 & 35.12 & 0.26 & 0.29 & 0.93 & 1.04 & 86\% & 90\% & 90\%\\

PKI  & 0.30 & 1.07 & 26.59 & 38.35 & 0.28 & 0.32 & 1.00 & 1.13 & 82\% & 88\% & 87\%\\

R    & 0.37 & 1.02 & 18.51 & 15.76 & 0.33 & 0.41 & 0.90 & 1.15 & 57\% & 58\% & 58\%\\

RAI  & 0.35 & 1.03 & 24.94 & 40.46 & 0.32 & 0.38 & 0.98 & 1.08 & 86\% & 76\% & 76\%\\

SLB  & 0.35 & 1.06 & 18.98 & 40.60 & 0.31 & 0.38 & 1.01 & 1.12 & 91\% & 88\% & 88\%\\

TE   & 0.21 & 1.00 & 16.18 & 24.28 & 0.18 & 0.24 & 0.92 & 1.09 & 70\% & 86\% & 86\%\\

TWC  & 0.37 & 1.04 & 17.70 & 15.96 & 0.33 & 0.42 & 0.91 & 1.16 & 72\% & 79\% & 79\%\\

WHR  & 0.78 & 1.18 & 9.24 & 11.54 & 0.61 & 0.94 & 0.98 & 1.38 & 44\% & 43\% & 42\%\\

WIN  & 5.81 & 1.60 & 16.09 & 11.70 & 5.11 & 6.52 & 1.33 & 1.87 & 28\% & 71\% & 71\%\\

WPI  & 0.27 & 0.92 & 19.33 & 28.99 & 0.24 & 0.30 & 0.86 & 0.98 & 78\% & 76\% & 76\%\\

XTO  & 0.31 & 1.04 & 30.85 & 39.51 & 0.29 & 0.33 & 0.98 & 1.09 & 89\% & 91\% & 91\%\\
\hline
Grand mean & 0.45 & 0.98 & 20.74 & 29.53 & 0.38 & 0.52 & 0.88 & 1.08 & 74\% & 75\% & 75\%\\
\hline
\end{tabular}
\end{center}
\vskip 12 pt
\scriptsize

\noindent Table 3 presents the results of regressions:
\vskip 6pt
$\log{\hat{\beta}_i}=\hat{\alpha_{L,i}}-\hat{\lambda}\log{AD_i}+\hat{\epsilon}_{L,i}$,

$\hat{\beta}_i=\hat{\alpha_{M,i}}+\frac{\hat{c}}{AD^{\hat{\lambda}}_i}+\hat{\epsilon}_{M,i}$,
\vskip 6pt
\noindent where $\hat{\beta}_i$ is the price impact coefficient for the $i$-th half-hour subsample and $AD_i$ is the average market depth for that subsample. These regressions were estimated for each of the 50 stocks, using 273 estimates of $\hat{\beta}_i$ for that stock, obtained from (\ref{simpleregr}). The second regression uses estimates $\hat{\lambda}$ obtained from the first regression. The t-statistics and the confidence intervals were computed using Newey-West standard errors. Confidence intervals are built with normal critical values. The last three columns provide three alternative fit measures - the $R^2$ of the linear regression (\ref{beta_depth_regr}), the squared correlation between $\hat{\beta}_i$ and fitted values $\hat{\hat{\beta}}_i=\frac{\hat{c}}{AD_i^{\hat{\lambda}}}$ and the squared correlation between $\hat{\beta}_i$ and $\hat{\hat{\beta}}^*_i=\frac{\hat{c}}{AD_i}$.
\normalsize

\newpage

\subsection{Intraday patterns}

\begin{figure}[h!]
\noindent
\begin{center}
{
\includegraphics[width=120mm]{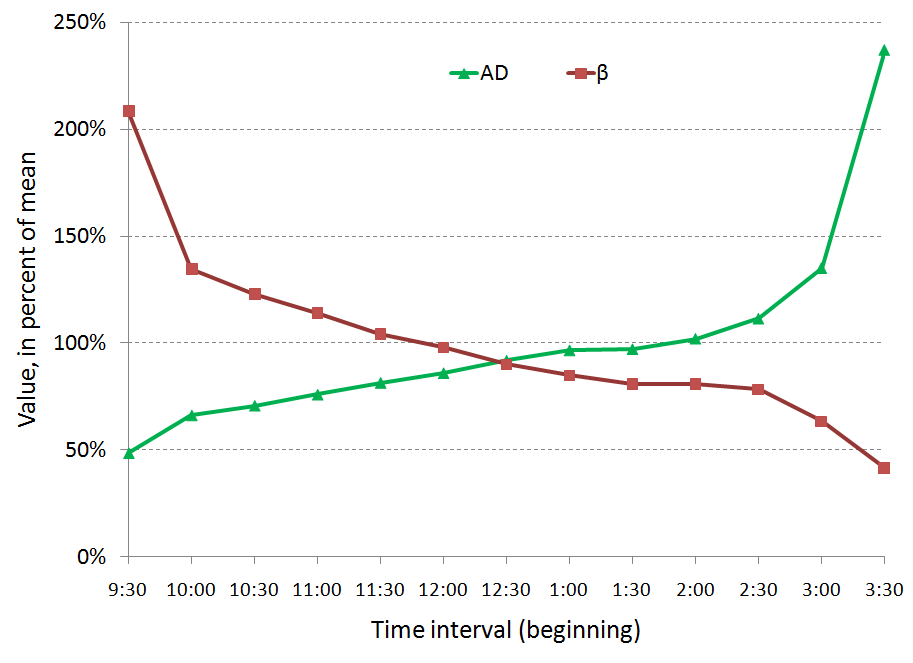}
}
\end{center}
\caption{Intraday patterns in the price impact coefficient $\hat{\beta}_i$ and the average depth $AD_i$.}
\label{int_betadepth}
\end{figure}

The link that we established between the price impact and the market depth has an important implication. Since the market depth follows a predictable pattern of intraday seasonality (\cite{ahn01}, \cite{lee93}), the price impact coefficient must also have a {\it predictable} intraday pattern. To demonstrate it, we averaged $\hat{\beta}_i$ for each stock and each half-hour interval across days, resulting in the intraday seasonality pattern for that stock, normalized these values by the average $\hat{\beta}_i$ of that stock and averaged the normalized seasonality patterns across stocks. The same procedure was repeated for $AD_i$ and the results are shown on Figure \ref{int_betadepth}.

Near the market open, depth is two times lower than it is on average, indicating that the order book is relatively shallow. In a shallow market, the incoming orders can easily affect the mid-price and the price impact coefficient is two times higher near the market open than on average. Moreover, price impact is five times higher at the market open compared to the market close.

The intraday pattern in price impact can be used to explain the intraday patterns in price volatility, observed by many researchers (\cite{ahn01}, \cite{andersen98}, \cite{hasbrouck91-2}, \cite{madhavan97}). Similarly to the price impact coefficient and the market depth, we computed the intraday patterns in variances of $\Delta P_k$ and $OFI_k$, using half-hour subsamples (indexed by $i$). Taking the variance on both sides of equation (\ref{simplespec}) demonstrates the link between $var[\Delta P_{k}]_i$, $var[OFI_{k}]_i$ and $\beta_i$:

\begin{equation}\label{var_relation}
var[\Delta P_{k}]_i=\beta^2_ivar[OFI_{k}]_i+var[\epsilon_{k}]_i
\end{equation}

\begin{figure}[ht]
\noindent
\begin{center}
{
\includegraphics[width=120mm]{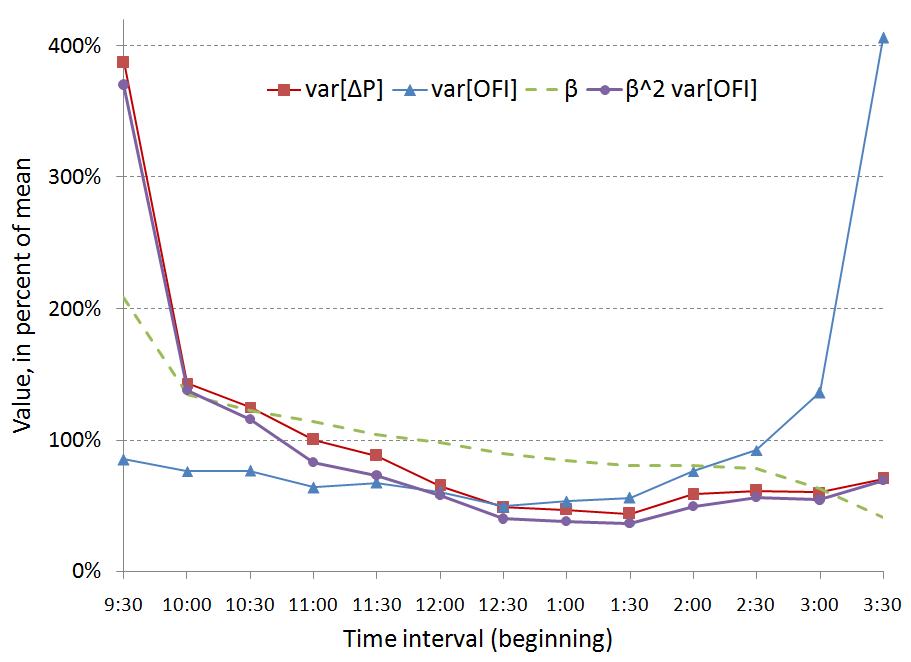}
}
\end{center}
\caption{Intraday seasonality in variances $var[\Delta P_{k}]_i$, $var[OFI_{k}]_i$, the price impact coefficient $\hat{\beta}_i$ and the expression $\beta^2_ivar[OFI_{k}]_i$.}
\label{INT}
\end{figure}

The average patterns are plotted on Figure \ref{INT}. Notice that the price volatility has a sharp peak near the market open, while the volatility of order flow imbalance has a peak near the market close. This peak is, however, offset by a low price impact, which gradually declines throughout the day. For the $i$-th half-hour interval, the equation (\ref{var_relation}) implies that $var[\Delta P_{k}]_i\approx\hat{\beta}^2_ivar[OFI_k]_i$ which is demonstrated on Figure \ref{INT}\footnote{$\hat{\beta}^2_ivar[OFI_k]_i$ was computed from the average patterns of $\hat{\beta}_i$ and $var[OFI_k]_i$}.

The intraday pattern in price variance was explained by Madhavan et al. \cite{madhavan97} in terms of a structural model. They argued that the volatility is higher in the morning because of the higher inflow of both public and private information. Similarly, Hasbrouck \cite{hasbrouck91-2} argued that the peak of price volatility at market open is mostly due to higher intensity of public information. Both studies agree that the impact of trades is larger in the morning. Our model contributes to this discussion by explaining the peak of price volatility using tangible quantities, rather than unobservable parameters. We also argue that the price impact of trades and the information asymmetry may be, in fact, two sides of the same coin.

First, we associate the higher volatility of order flow imbalance at market the open and close with a higher rate of trading, that is, higher inflow of public and private information. Second, if the bid-ask spread is small (it is mostly equal to 1 cent in our data), limit order traders may avoid being ``picked off" only by lowering the number of submitted orders, reducing the depth. Therefore, if limit order traders are aware of information asymmetry in the morning, the low depth may simply indicate this asymmetry. In our model,  low depth also { implies} a higher price impact, making the information advantages harder to realize at the market open.

\newpage

\section{Price impact of trades}\label{volume.sec}
\subsection{Trade imbalance vs order flow imbalance}
The previous section discussed the linear relation between price changes and $OFI_{k}$ - our measure of supply/demand imbalance. However, little has been said about trade imbalances, which are widely used in the academic literature \cite{chordia08,hasbrouck91,hasbrouck01,hopman07,kempf99,plerou02} and in practice \cite{torre97}. The aim of this section is to compare the price impact of trades and order flow imbalance and show that the (nonlinear) price impact of trade volume may be derived from our linear model for the price impact of order flow.

For convenience we will call `buy trade' a transaction initiated by a market buy order and `sell trade' a transaction initiated by a market sell order.
We define the
 {\it trade imbalance} during a time interval $[t_{k-1},t_k]$ as the difference between volumes of buy and sell trades during that interval:
$$TI_{k} =\sum_{n=N(t_{k-1})+1}^{N(t_{k})}{b_n}-\sum_{n=N(t_{k-1})+1}^{N(t_{k})}{s_n},$$
\noindent Here, $b_n$ is the size of a buyer-initiated trade that occurs at the $n$-th quote; $b_n=0$ if no buy trade occurs at that quote. Similarly, $s_n$ is the size of a sell trade that occurs at the $n$-th quote or zero. The procedure that matches trades with quotes and classifies them as buys or sells is described in the Appendix.

To compare the explanatory power of trade and order flow imbalances with respect to price changes, we perform the following  regressions:
\begin{subequations}
\begin{gather}\label{dp_imb_regr1}
\Delta P_{k}=\hat{\alpha}_i+\hat{\beta}_iOFI_{k}+\hat{\epsilon}_{k}
\end{gather}
\begin{gather}\label{dp_imb_regr2}
\Delta P_{k}=\hat{\alpha}_{T,i}+\hat{\beta}_{T,i}TI_k+\hat{\epsilon}_{T,k}
\end{gather}
\begin{gather}\label{dp_imb_regr3}
\Delta P_{k}=\hat{\alpha}_{D,i}+\hat{\theta}_{O,i}OFI_k+\hat{\theta}_{T,i}TI_k+\hat{\epsilon}_{D,k}
\end{gather}
\end{subequations}

\noindent The regressions are estimated separately for every half-hour subsample of data (indexed by $i$). If the effect of trades is included in the order flow imbalance, the coefficients $\hat{\theta}_{T,i}$ in (\ref{dp_imb_regr3}) must be indistinguishable from zero. We note that regressions (\ref{dp_imb_regr1}-\ref{dp_imb_regr3}) contain only the linear terms, because we found no evidence of non-linear price impacts in our data (for neither $OFI_k$ nor $TI_k$). The average results of these regressions are presented in Panel A of Table 4. Clearly, when $OFI_k$ and $TI_k$ are taken individually, each of them has a statistically significant influence on price changes. Comparing the two we observe that $OFI_k$ explains price changes better than $TI_k$ - the average $R^2$ for order flow imbalance is 65\% compared to 32\% for the trade imbalance. When two variables are used together to explain price changes, the dependence on trade imbalance becomes questionable. The average t-statistic of $TI_k$ decreases by a factor of four and the coefficients $\hat{\theta}_{T,i}$ are statistically significant in only 31\% of subsamples. However, the dependence on $OFI_k$ remains convincingly strong.

Our findings show that:
\begin{enumerate}
\item The order flow imbalance $OFI_k$ explains price movements better than the imbalance of trades.
\item The effect of trade imbalance is adequately included in $OFI_k$, a more general measure of supply/demand imbalance.
\end{enumerate}

\newpage

\addtolength{\oddsidemargin}{-.3in}
\addtolength{\evensidemargin}{-.3in}

\begin{center}
{Table 4. Comparison of order flow imbalance and trade imbalance.}
\vskip 6pt
\scriptsize
\begin{tabular}{|l|r r r r|r r r r|r r r r r r|}
\hline
\multicolumn{15}{|c|}{Panel A: Detailed results for changes in mid prices}\\
\hline
\multirow{2}{*}{Ticker} & \multicolumn{4}{|c|}{Order flow imbalance} & \multicolumn{4}{|c|}{Trade imbalance} & \multicolumn{6}{|c|}{Both covariates}\\
\cline{2-15}
 & $R^2$ & $t(\hat{\beta})$ & $\{\beta\neq 0\}$ & $F$ & $R^2$ & $t(\hat{\beta}_T)$ & $\{\beta_T\neq 0\}$ & $F$ & $R^2$ & $t(\hat{\theta}_O)$ & $t(\hat{\theta}_T)$ & $\{\theta_O\neq 0\}$ & $\{\theta_T\neq 0\}$ & $F$\\
\hline

AMD  & 64\% & 9.96 & 98\% & 382 & 39\% & 4.15 & 86\% & 140 & 67\% & 6.49 & 1.26 & 93\% & 34\% & 214\\

APOL & 63\% & 10.32 & 91\% & 396 & 30\% & 4.14 & 84\% & 83 & 66\% & 8.00 & 1.09 & 89\% & 26\% & 211\\

AXP  & 69\% & 13.87 & 100\% & 449 & 34\% & 4.72 & 83\% & 101 & 71\% & 10.05 & 1.50 & 100\% & 44\% & 241\\

AZO  & 47\% & 6.39 & 97\% & 179 & 30\% & 4.09 & 90\% & 87 & 54\% & 5.02 & 2.34 & 96\% & 68\% & 118\\

BAC  & 79\% & 18.36 & 100\% & 774 & 45\% & 6.31 & 96\% & 157 & 80\% & 12.55 & 0.72 & 99\% & 19\% & 397\\

BDX  & 63\% & 10.08 & 100\% & 362 & 28\% & 4.02 & 82\% & 79 & 65\% & 7.88 & 1.23 & 97\% & 34\% & 195\\

BK   & 74\% & 14.97 & 100\% & 610 & 36\% & 4.58 & 81\% & 117 & 75\% & 10.68 & 0.68 & 99\% & 17\% & 313\\

BSX  & 58\% & 6.12 & 81\% & 338 & 31\% & 2.57 & 54\% & 106 & 62\% & 4.51 & 0.57 & 73\% & 12\% & 189\\

BTU  & 72\% & 14.51 & 100\% & 527 & 35\% & 5.21 & 88\% & 103 & 74\% & 10.90 & 1.31 & 99\% & 32\% & 277\\

CAT  & 71\% & 14.85 & 99\% & 498 & 33\% & 5.01 & 86\% & 94 & 72\% & 11.27 & 1.28 & 99\% & 38\% & 262\\

CB   & 64\% & 11.97 & 100\% & 378 & 33\% & 4.66 & 88\% & 102 & 66\% & 8.42 & 1.34 & 99\% & 37\% & 202\\

CCL  & 70\% & 13.88 & 99\% & 478 & 32\% & 4.55 & 85\% & 93 & 71\% & 10.50 & 0.98 & 99\% & 26\% & 247\\

CINF & 70\% & 10.73 & 98\% & 552 & 39\% & 4.26 & 87\% & 141 & 72\% & 7.17 & 1.01 & 96\% & 27\% & 297\\

CME  & 35\% & 4.98 & 94\% & 112 & 24\% & 3.39 & 75\% & 63 & 44\% & 4.10 & 2.18 & 92\% & 59\% & 78\\

COH  & 69\% & 12.75 & 100\% & 457 & 29\% & 3.91 & 82\% & 80 & 70\% & 10.06 & 0.87 & 100\% & 22\% & 238\\

COP  & 68\% & 12.50 & 100\% & 450 & 35\% & 4.92 & 84\% & 107 & 70\% & 9.19 & 1.42 & 100\% & 40\% & 240\\

CVH  & 65\% & 10.83 & 99\% & 418 & 35\% & 4.10 & 84\% & 114 & 67\% & 7.30 & 1.01 & 97\% & 25\% & 222\\

DNR  & 69\% & 12.76 & 99\% & 471 & 32\% & 3.98 & 81\% & 101 & 70\% & 9.29 & 1.01 & 97\% & 24\% & 246\\

DVN  & 65\% & 11.48 & 97\% & 414 & 33\% & 4.83 & 88\% & 96 & 68\% & 8.58 & 1.70 & 93\% & 48\% & 226\\

EFX  & 56\% & 8.71 & 98\% & 289 & 31\% & 3.72 & 80\% & 101 & 60\% & 6.21 & 1.64 & 96\% & 43\% & 167\\

ETN  & 65\% & 10.51 & 98\% & 389 & 25\% & 3.59 & 71\% & 69 & 67\% & 8.66 & 1.04 & 98\% & 29\% & 209\\

FISV & 63\% & 10.42 & 100\% & 380 & 28\% & 3.79 & 81\% & 79 & 65\% & 8.12 & 0.88 & 100\% & 25\% & 201\\

HAS  & 67\% & 11.45 & 100\% & 427 & 32\% & 4.04 & 84\% & 97 & 68\% & 8.53 & 0.89 & 100\% & 24\% & 223\\

HCP  & 67\% & 13.60 & 100\% & 417 & 31\% & 4.43 & 82\% & 91 & 68\% & 10.01 & 1.05 & 100\% & 32\% & 217\\

HOT  & 68\% & 12.64 & 99\% & 438 & 27\% & 3.86 & 77\% & 74 & 70\% & 9.94 & 1.17 & 99\% & 29\% & 231\\

KSS  & 71\% & 13.82 & 98\% & 525 & 31\% & 4.43 & 81\% & 91 & 72\% & 10.83 & 0.94 & 97\% & 25\% & 274\\

LLL  & 67\% & 11.76 & 96\% & 485 & 36\% & 5.07 & 90\% & 117 & 70\% & 8.58 & 1.63 & 94\% & 44\% & 270\\

LMT  & 72\% & 13.58 & 100\% & 516 & 35\% & 4.89 & 90\% & 105 & 73\% & 10.19 & 1.50 & 99\% & 40\% & 277\\

M    & 75\% & 15.82 & 100\% & 640 & 35\% & 4.41 & 84\% & 108 & 76\% & 11.38 & 0.97 & 100\% & 26\% & 330\\

MAR  & 71\% & 14.61 & 100\% & 498 & 34\% & 4.77 & 89\% & 105 & 72\% & 10.45 & 1.05 & 100\% & 27\% & 258\\

MFE  & 68\% & 12.72 & 100\% & 463 & 31\% & 4.17 & 82\% & 93 & 69\% & 9.06 & 0.73 & 99\% & 18\% & 239\\

MHP  & 68\% & 11.62 & 99\% & 489 & 31\% & 3.85 & 84\% & 96 & 70\% & 8.92 & 0.77 & 98\% & 19\% & 257\\

MHS  & 66\% & 11.70 & 99\% & 414 & 28\% & 4.03 & 77\% & 80 & 68\% & 9.10 & 1.11 & 99\% & 27\% & 218\\

MRK  & 69\% & 12.53 & 100\% & 451 & 31\% & 4.08 & 82\% & 93 & 70\% & 9.20 & 0.76 & 100\% & 20\% & 235\\

MRO  & 69\% & 13.67 & 100\% & 465 & 35\% & 4.66 & 89\% & 104 & 70\% & 9.73 & 0.91 & 100\% & 24\% & 241\\

MWV  & 68\% & 11.79 & 100\% & 452 & 34\% & 4.37 & 86\% & 102 & 69\% & 8.63 & 0.80 & 100\% & 24\% & 237\\

NEM  & 71\% & 13.81 & 100\% & 490 & 34\% & 4.99 & 81\% & 100 & 72\% & 10.24 & 1.53 & 99\% & 43\% & 260\\

OMC  & 65\% & 11.88 & 99\% & 411 & 30\% & 4.14 & 85\% & 88 & 67\% & 8.99 & 0.96 & 99\% & 24\% & 216\\

PCS  & 53\% & 5.21 & 79\% & 297 & 35\% & 2.68 & 59\% & 169 & 58\% & 3.44 & 0.86 & 71\% & 20\% & 195\\

PHM  & 66\% & 10.33 & 98\% & 416 & 35\% & 3.87 & 84\% & 115 & 68\% & 7.28 & 0.95 & 93\% & 29\% & 224\\

PKI  & 53\% & 7.25 & 94\% & 263 & 28\% & 3.03 & 70\% & 89 & 57\% & 5.39 & 1.24 & 88\% & 32\% & 148\\

R    & 63\% & 10.14 & 98\% & 352 & 27\% & 3.92 & 86\% & 71 & 65\% & 8.07 & 1.20 & 97\% & 30\% & 188\\

RAI  & 66\% & 10.40 & 100\% & 422 & 36\% & 4.67 & 89\% & 111 & 68\% & 7.52 & 1.11 & 99\% & 31\% & 224\\

SLB  & 76\% & 16.76 & 100\% & 644 & 32\% & 4.54 & 79\% & 94 & 77\% & 13.02 & 1.24 & 100\% & 36\% & 336\\

TE   & 54\% & 6.66 & 86\% & 301 & 37\% & 3.27 & 67\% & 175 & 60\% & 4.34 & 1.32 & 79\% & 29\% & 200\\

TWC  & 64\% & 11.80 & 99\% & 377 & 31\% & 4.26 & 77\% & 93 & 66\% & 8.46 & 1.34 & 99\% & 37\% & 201\\

WHR  & 65\% & 10.26 & 97\% & 394 & 29\% & 4.29 & 88\% & 85 & 67\% & 8.17 & 1.43 & 96\% & 39\% & 217\\

WIN  & 44\% & 3.12 & 60\% & 243 & 41\% & 2.68 & 54\% & 249 & 58\% & 1.78 & 1.39 & 42\% & 29\% & 206\\

WPI  & 66\% & 10.47 & 98\% & 437 & 32\% & 3.91 & 83\% & 100 & 68\% & 7.82 & 1.05 & 97\% & 30\% & 232\\

XTO  & 65\% & 13.28 & 100\% & 399 & 21\% & 3.05 & 63\% & 54 & 66\% & 10.72 & 1.05 & 100\% & 27\% & 209\\
\hline
Grand mean & 65\% & 11.47 & 97\% & 429 & 32\% & 4.18 & 81\% & 103 & 67\% & 8.49 & 1.16 & 95\% & 31\% & 231\\
\hline
\multicolumn{15}{|c|}{Panel B: Average results for changes in transaction prices}\\
\hline

$L=2$ trades & 14\% & 15.74 & 98\% & 464 & 1\% & 2.69 & 63\% & 26 & 15\% & 14.17 & -2.58 & 98\% & 54\% & 245\\

$L=5$ trades & 38\% & 19.42 & 98\% & 753 & 8\% & 4.50 & 75\% & 113 & 39\% & 16.85 & -0.20 & 98\%  & 9\% & 379\\

$L=10$ trades & 51\% & 17.78 & 98\% & 655 & 13\% & 4.55 & 75\% & 100 & 51\% & 14.97 & 0.57 & 98\% &  9\% & 329\\

\hline

\end{tabular}
\end{center}
\vskip 2pt
\tiny
\noindent Table 4 presents the average results of regressions:
\vskip 2pt
$\Delta P_{k}=\hat{\alpha}_i + \hat{\beta}_iOFI_{k}+\hat{\epsilon}_{k}$,

$\Delta P_{k}=\hat{\alpha}_{T,i} + \hat{\beta}_{T,i}TI_{k}+\hat{\epsilon}_{T,k}$,

$\Delta P_{k}=\hat{\alpha}_{D,i} + \hat{\theta}_{O,i}OFI_{k} + \hat{\theta}_{T,i}TI_{k} +\hat{\epsilon}_{D,k}$,
\vskip 2pt
\noindent where $\Delta P_{k}$ are the 10-second mid-price changes (Panel A) or changes in trade prices between $L$ trades (Panel B), $OFI_{k}$ are the contemporaneous order flow imbalances and $TI_{k}$ are the contemporaneous trade imbalances. For Panel A, these regressions were estimated using 273 half-hour subsamples (indexed by $i$) for each stock and their outputs were averaged across subsamples. Each subsample typically contains about 180 observations (indexed by $k$). For Panel B, data was pooled across half-hour subsamples, resulting in 13 subsamples for each stock. The t-statistics were computed using White's standard errors. For each of three regressions, Table 4 reports the average $R^2$, the average t-statistic of the coefficient(s), the percentage of samples where the coefficient(s) passed the z-test at the 5\% significance level and the  F-statistic of the regression. The outputs for Panel B were averaged across stocks.

\newpage
\normalsize

\addtolength{\oddsidemargin}{.3in}
\addtolength{\evensidemargin}{.3in}

As a robustness check, we repeated regressions (\ref{dp_imb_regr1}-\ref{dp_imb_regr3}) with differences between transaction prices $P^t_{k}$ instead of differences in mid-prices $P_{k}$. This time price differences were computed in {\it trade time} as $\Delta_L P^t_{k}=P^t_{k}-P^t_{k-L}$ for $L$ trades. The average results across five stocks, picked at random\footnote{The stocks tickers were BDX, CB, MHS, PHM and PKI. We computed price changes for $L=2,5,10$ trades to mitigate the possible issues with trade and quote alignment in the TAQ data and we correspondingly computed order flow imbalances and trade imbalances during the time intervals between 2, 5 or 10 consecutive trades. To ensure that there is an ample amount of data for each regression, we pooled data across days for each stock and each time interval.} are presented in Panel B of Table 4. Our findings for transaction prices are essentially the same as for mid prices - $OFI_k$ explains price changes better than $TI_k$. Moreover, the effect of trades on prices seems to be captured by the order flow imbalance. The variable $TI_k$ becomes statistically insignificant when used together with $OFI_k$ in the regression and the increase in $R^2$ from adding $TI_k$ as an extra regressor is not economically significant. 

Interestingly, we found that the relation between $\Delta_L P^t_{k}$ and $OFI_k$ is concave in some samples, and similarly for $\Delta_L P^t_{k}$ and $TI_k$. We estimated regressions (\ref{dp_imb_regr1}) and (\ref{dp_imb_regr2}) for transaction price changes with additional quadratic terms $OFI_{k}|OFI_{k}|$ (respectively, $TI_{k}|TI_{k}|$) and found that they are significant in nearly half of the samples with t-statistics of -2.8 on average (-2.3 for $TI_{k}|TI_{k}|$). Sampling data at special times (trade times) may introduce biases to the right side of the regression. One possible explanation is that traders submit their orders when they expect their impact to be minimal, leading to a concave (sublinear) impact. Supporting this idea of sampling biases, we found that when mid-prices are sampled at trade times, the price impact of $OFI_k$ is again concave in some samples. On another hand, when we regressed last trade prices sampled at 1-minute frequency on $OFI_k$, we observed the concave price impact once again. This suggests that using either trade times or trade prices may lead to non-linear price impact. However the quadratic term in our regressions is insignificant in about half of the samples and marginally significant in the the other half of the data.

\subsection{Does volume move the prices?}
The relation between price changes and volume is empirically confirmed by many authors (see \cite{karpoff87} for a review). Recently, traded volume became an important metric for order execution algorithms - these algorithms often attempt to match a certain percentage of the total traded volume to reduce the price impact. However, it remains unclear whether the traded volume truly determines the magnitude of price moves and whether it is a good metric for price impact. Casting doubt on this assertion, Jones et al. \cite{jones94} showed that the relation between the daily volatility and the daily volume is essentially due to the number of trades and not the volume per se (also see \cite{Chan2000} for the discussion).

We extend this result in two ways. First, we show that even when prices are driven by order flow imbalance, an apparent (concave) dependence on traded volume may emerge as an artifact due to data aggregation. Second, we empirically confirm that the price-volume relation is an indirect one - it becomes statistically insignificant after accounting for the order flow imbalance.

The  volume traded during a time interval $[t_{k-1},t_k]$ is:
$$VOL_{k} =\sum_{n=N(t_{k-1})+1}^{N(t_{k})}{b_n}+\sum_{n=N(t_{k-1})+1}^{N(t_{k})}{s_n}=\sum_{n=N(t_{k-1})+1}^{N(t_{k})}{w_n},$$
\noindent where $w_n=b_n+s_n$ is the size of any trade (either buy or sell) if it occurs at the $n$-th quote or zero otherwise. Comparing this definition with the definition of $OFI_k$ we note that both quantities are sums of random variables. As the aggregation window $[t_{k-1},t_k]$ becomes progressively larger, the  behavior of these sums (under certain assumptions) will be governed by the Law of Large Numbers and the Central Limit Theorem. We consider a time interval $[0,T)$ and denote by $N(T)$ the number of order book events during that time interval. We also denote by $OFI(T)$ and $VOL(T)$, respectively, the order flow imbalance and the traded volume during $[0,T)$. The following proposition shows a link between $OFI(T)$ and $VOL(T)$ as $T$ grows.

\begin{myprop}\label{sqrt_vol_prop}
Assume that
\begin{enumerate}
\item Order book events accumulate over time at some average rate $\Lambda$:
$\frac{N(T)}{T}\rightarrow \Lambda, ~as~ T\rightarrow\infty$

\item $\{e_i\}_{i=1}^{\infty}$ are i.i.d. random variables with a finite variance $\sigma^2$,

\item $\{w_i\}_{i=1}^{\infty}$ are i.i.d. random variables with a finite mean $\mu\pi$,
where $\pi$ is the proportion of order book events that correspond to trades and $\mu$ is the mean trade size.
\end{enumerate}
\begin{equation}\label{limit_V}
{\rm Then}\qquad\frac{\sqrt{\mu\pi}}{\sigma}\frac{OFI(T)}{\sqrt{VOL(T)}}\Rightarrow \xi,~as~ T\rightarrow\infty
\end{equation}
where $\xi\sim N(0,1)$ is a standard normal random variable and $\Rightarrow$ denotes convergence in distribution.
\end{myprop}
\textbf{Proof:}
First, we apply the law of large numbers to the traded volume. Assumption (1) ensures that $N(T)\rightarrow\infty$ as $T\rightarrow\infty$:
\begin{equation}\label{LLN_V}
\frac{VOL(T)}{N(T)}=\frac{\sum_{i=1}^{N(T)}w_i}{N(T)}\rightarrow \mu\pi, w.p. 1, ~as~ T\rightarrow\infty,
\end{equation}
Second, event contributions $e_i$ have a finite variance $\sigma^2$ and, under our assumptions, we can apply the classical central limit theorem to the order flow imbalance:
\begin{equation}\label{CLT_I}
\frac{OFI(T)}{\sigma\sqrt{N(T)}}\equiv\frac{\sum_{i=1}^{N(T)}{e_i}}{\sigma\sqrt{N(T)}}\Rightarrow \xi, ~as~ T\rightarrow\infty,
\end{equation}
where $\xi\sim N(0,1)$ is a standard normal random variable. Although the denominator $\sigma\sqrt{N(T)}$ is random, it goes to infinity by assumption (1) and Anscombe's lemma ensures that we can use such a normalization in the central limit theorem \cite[Lemma 2.5.8]{embrechts97}.
Since the square root function is continuous, the convergence in (\ref{LLN_V}) takes place almost-surely and
  the limit in (\ref{LLN_V}) is deterministic, we can combine (\ref{LLN_V}) and (\ref{CLT_I}) in the following way:
\begin{equation}
\frac{\sqrt{\mu\pi}}{\sigma}\frac{OFI(T)}{\sqrt{VOL(T)}}\equiv
\frac{\frac{\sum_{i=1}^{N(T)}{e_i}}{\sigma\sqrt{N(T)}}}{\sqrt{\frac{\sum_{i=1}^{N(T)}{w_i}}{\mu\pi (N(T))}}}\Rightarrow \xi,~as~ T\rightarrow\infty
\end{equation}
$\blacksquare$

If the time interval $[0,T)$ includes a large enough number of order book events and trades, the above limit argument implies a noisy scaling relation between order flow imbalance and the square root of traded volume:
\begin{equation}\label{I_V}
OFI(T)=\xi\frac{\sigma}{\sqrt{\mu\pi}}\sqrt{VOL(T)},
\end{equation}

\noindent where $\mu,\pi$ and $\sigma$ are constants and $\xi\sim N(0,1)$. Now, assume that it holds not just for the first interval, but for every time interval $[t_{k-1}, t_k)$ of large enough length $\Delta t$, regardless of its index $k$. Then, (\ref{I_V}) can be substituted into our model (\ref{simplespec}), to yield:

\begin{equation}\label{sqrt_1}
\Delta P_{k}= \theta_{k}\sqrt{VOL_{k}}+\epsilon_{k},
\end{equation}
\noindent where $\theta_{k}=\beta_{i}\xi_{k}\frac{\sigma}{\sqrt{\mu\pi}}$ is a slope coefficient and $\xi_{k}\sim N(0,1)$ is a noise term due to scaling. Due to the scaling approximation, the slope $\theta_{k}$ in (\ref{sqrt_1}) is a random normal variable: $\theta_{k}\sim N(0,\beta^2_{i}\frac{\sigma^2}{\mu\pi})$. For every time interval $[t_{k-1}, t_k)$ the ratio $\frac{\sqrt{\mu\pi}}{\sigma}\frac{OFI_{k}}{\sqrt{VOL_k}}$ is a different draw from the $N(0,1)$ distribution, leading to a different $\theta_{k}$ in each case. This additional randomness makes this model considerably less robust than (\ref{simplespec}) and we do not recommend to use it.

Equation (\ref{sqrt_1}) shows that even if prices are  driven by the order flow imbalance (i.e. even if $\epsilon_{k}=0~\forall k$), there will be a noisy square-root relation between the price changes and the traded volume. However, if the assumptions of Proposition \ref{sqrt_vol_prop} do not hold (e.g. $\{e_i\}_{i=1}^{\infty}$ are strongly dependent or have infinite variance), the price-volume relation may have a different exponent. A variety of exponents $0<H<1$ have been observed in the relation between prices changes and trade sizes  \cite{bouchaud08}, suggesting the following model:
\begin{equation}\label{sqrt_2}
\Delta P_{k}= \theta_{k}VOL^H_{k}+\epsilon_{k},
\end{equation}
To estimate the exponent $H$, we put $\epsilon_{k}=0$ and $\theta_{k}=\bar{\theta}_i\xi_{k}$  in (\ref{sqrt_2}) and fit a logarithmic regression to every half-hour subsample, indexed by $i$:
\begin{equation}\label{sqrt_regr}
\log{|\Delta P_{t_k}|}= \log{\hat{\bar{\theta}}_i}+\hat{H}_i\log{VOL_{k}}+\log{\hat{\xi}_{k}}
\end{equation}
Based on Proposition \ref{sqrt_vol_prop}, we expect the price-volume relation to be indirect (i.e. come through $OFI_k$) and noisy. To empirically confirm this, we compare the following three regressions:
\begin{subequations}
\begin{gather}\label{dp_vol_regr1}
|\Delta P_{k}|=\hat{\alpha}_{O,i}+\hat{\beta}_{O,i}|OFI_{k}|+\hat{\epsilon}_{O,k}
\end{gather}
\begin{gather}\label{dp_vol_regr2}
|\Delta P_{k}|=\hat{\alpha}_{V,i}+\hat{\beta}_{V,i}VOL^{\hat{H}_i}_{k}+\hat{\epsilon}_{V,k}
\end{gather}
\begin{gather}\label{dp_vol_regr3}
|\Delta P_{k}|=\hat{\alpha}_{W,i}+\hat{\phi}_{O,i}|OFI_{k}|+\hat{\phi}_{V,i}VOL^{\hat{H}_i}_{k}+\hat{\epsilon}_{W,k}
\end{gather}
\end{subequations}

\noindent These regressions are estimated for every half-hour subsample with the exponents $\hat{H}_i$ pre-estimated by (\ref{sqrt_regr}). The averages of $\hat{H}_i$ and their standard deviation for each stock are presented on the left panel in Table 5. The exponent varies considerably across stocks and time, but is generally below 1/2 in our data.
The average results of regressions (\ref{dp_vol_regr1}-\ref{dp_vol_regr3}) for each stock are presented on the middle and right panels. We observe that $|OFI_{k}|$ explains the magnitude of price moves better than $VOL^{\hat{H}_i}_{k}$. Although both variables appear to be statistically significant when taken individually, only $|OFI_{k}|$ remains significant in the multiple regression. Thus, the dependence between the magnitude of price moves and the traded volume is mostly due to correlation between $VOL_{k}$ and $|OFI_{k}|$. Interestingly, the number of trades variable (suggested in \cite{jones94}) is also statistically significant on a stand-alone basis, but becomes insignificant when added to (\ref{dp_vol_regr3}) as a third variable.

\newpage

\section{Conclusion}\label{conclusion.sec}

We have introduced {\it order flow imbalance}, a variable that cumulates the sizes of order book events, treating the contributions of market, limit and cancel orders equally, and
provided empirical and theoretical evidence for a linear relation between high-frequency  price changes and order flow imbalance for individual stocks. We have shown that this linear model is robust across stocks and the impact coefficient is inversely proportional to market depth. These relations suggest that prices respond to changes in the supply and demand for shares at the best quotes, and that the impact coefficient fluctuates with the amount of liquidity provision, or depth, in the market. Moreover, we have demonstrated that order flow imbalance is a stronger driver of high-frequency price changes than standard measures of trade imbalance. Trades seem to carry little to no information about price changes after the simultaneous order flow imbalance is taken into account. If trades do not help to explain price changes after controlling for the order flow imbalance, it is highly possible that the relation between price changes and traded volume simply capture the noisy scaling relation between these variables.

Overall, these findings seem to give an intuitive picture of the price impact of order book events, which is somewhat simpler than the one conveyed by previous studies.
\newpage

\addtolength{\oddsidemargin}{-.5in}
\addtolength{\evensidemargin}{.5in}

\begin{center}
{Table 5. Comparison of traded volume and order flow imbalance.}
\scriptsize
\vskip 6pt
\begin{tabular}{|l|r r|r r r r|r r r r|r r r r r r|}
\hline
\multirow{2}{*}{Ticker} & Avg  & Stdev & \multicolumn{4}{|c|}{Order flow imbalance} & \multicolumn{4}{|c|}{Traded volume} & \multicolumn{6}{|c|}{Both covariates}\\
\cline{4-17}
 & $\hat{H}$ & $\hat{H}$ & $R^2$ & $t(\hat{\beta}_O)$ & ${\beta_O\neq 0}$ & $F$ & $R^2$ & $t(\hat{\beta}_V)$ & ${\beta_V\neq 0}$ & $F$ & $R^2$ & $t(\hat{\phi}_O)$ & $t(\hat{\phi}_V)$ & ${\phi_O\neq 0}$ & ${\phi_V\neq 0}$ & $F$\\
\hline

AMD  & 0.06 & 0.08 & 63\% & 10.3 & 99\% & 356 & 14\% & 4.5 & 83\% & 34 & 63\% & 9.4 & 1.1 & 99\% & 35\% & 182\\

APOL & 0.24 & 0.08 & 53\% & 8.3 & 90\% & 258 & 25\% & 6.8 & 99\% & 63 & 57\% & 6.9 & 2.9 & 89\% & 84\% & 144\\

AXP  & 0.16 & 0.08 & 55\% & 10.5 & 100\% & 249 & 20\% & 6.6 & 100\% & 48 & 57\% & 9.0 & 2.8 & 100\% & 81\% & 133\\

AZO  & 0.43 & 0.22 & 39\% & 5.5 & 96\% & 131 & 32\% & 5.3 & 100\% & 93 & 50\% & 4.3 & 3.6 & 94\% & 96\% & 98\\

BAC  & 0.09 & 0.08 & 73\% & 16.3 & 100\% & 560 & 24\% & 5.6 & 83\% & 61 & 74\% & 13.9 & 1.2 & 96\% & 35\% & 285\\

BDX  & 0.26 & 0.10 & 55\% & 8.4 & 99\% & 261 & 27\% & 6.3 & 100\% & 71 & 58\% & 6.7 & 2.9 & 98\% & 84\% & 147\\

BK   & 0.11 & 0.07 & 68\% & 13.1 & 100\% & 437 & 19\% & 6.6 & 97\% & 46 & 68\% & 11.5 & 2.0 & 99\% & 58\% & 225\\

BSX  & -0.17 & 2.41 & 68\% & 8.4 & 100\% & 486 & 14\% & 3.3 & 95\% & 33 & 69\% & 8.0 & 0.1 & 97\% & 12\% & 246\\

BTU  & 0.24 & 0.07 & 58\% & 10.5 & 99\% & 283 & 23\% & 6.8 & 99\% & 57 & 60\% & 8.9 & 2.4 & 99\% & 78\% & 151\\

CAT  & 0.22 & 0.07 & 56\% & 10.4 & 98\% & 250 & 19\% & 6.0 & 98\% & 44 & 57\% & 8.9 & 2.1 & 98\% & 63\% & 131\\

CB   & 0.19 & 0.09 & 56\% & 10.1 & 99\% & 261 & 23\% & 6.4 & 99\% & 58 & 58\% & 8.2 & 2.6 & 99\% & 74\% & 141\\

CCL  & 0.14 & 0.07 & 60\% & 11.3 & 100\% & 309 & 19\% & 6.6 & 99\% & 45 & 62\% & 9.9 & 2.4 & 99\% & 74\% & 162\\

CINF & 0.13 & 0.12 & 67\% & 10.6 & 99\% & 505 & 30\% & 6.1 & 98\% & 85 & 69\% & 8.7 & 2.0 & 99\% & 55\% & 268\\

CME  & 0.49 & 0.24 & 28\% & 4.1 & 94\% & 78 & 30\% & 4.8 & 99\% & 83 & 42\% & 3.2 & 3.6 & 86\% & 94\% & 71\\

COH  & 0.19 & 0.07 & 60\% & 10.4 & 99\% & 299 & 22\% & 6.5 & 99\% & 52 & 61\% & 8.9 & 2.2 & 98\% & 69\% & 157\\

COP  & 0.16 & 0.07 & 56\% & 9.8 & 100\% & 277 & 20\% & 6.0 & 96\% & 49 & 58\% & 8.4 & 2.4 & 100\% & 70\% & 145\\

CVH  & 0.18 & 0.10 & 62\% & 10.2 & 100\% & 352 & 27\% & 5.9 & 99\% & 72 & 64\% & 8.2 & 2.2 & 100\% & 70\% & 189\\

DNR  & 0.08 & 0.07 & 64\% & 12.0 & 99\% & 376 & 17\% & 6.3 & 95\% & 38 & 65\% & 10.7 & 1.8 & 99\% & 55\% & 193\\

DVN  & 0.26 & 0.07 & 52\% & 8.6 & 93\% & 236 & 24\% & 6.7 & 100\% & 59 & 55\% & 7.1 & 2.9 & 91\% & 81\% & 131\\

EFX  & 0.20 & 0.11 & 52\% & 8.1 & 99\% & 241 & 26\% & 5.4 & 99\% & 69 & 56\% & 6.4 & 2.7 & 97\% & 75\% & 137\\

ETN  & 0.26 & 0.10 & 55\% & 8.2 & 97\% & 252 & 27\% & 6.4 & 99\% & 70 & 58\% & 6.8 & 2.9 & 96\% & 83\% & 142\\

FISV & 0.19 & 0.11 & 57\% & 9.1 & 100\% & 284 & 25\% & 5.9 & 100\% & 65 & 59\% & 7.3 & 2.2 & 99\% & 66\% & 153\\

HAS  & 0.20 & 0.09 & 61\% & 10.1 & 100\% & 328 & 26\% & 6.2 & 100\% & 67 & 63\% & 8.2 & 2.3 & 100\% & 73\% & 175\\

HCP  & 0.14 & 0.07 & 57\% & 11.1 & 100\% & 268 & 21\% & 7.0 & 99\% & 50 & 59\% & 9.3 & 2.7 & 100\% & 79\% & 143\\

HOT  & 0.23 & 0.08 & 57\% & 9.7 & 98\% & 263 & 24\% & 6.9 & 100\% & 60 & 60\% & 8.2 & 3.0 & 98\% & 85\% & 145\\

KSS  & 0.24 & 0.08 & 60\% & 10.8 & 97\% & 318 & 25\% & 6.6 & 99\% & 61 & 62\% & 9.0 & 2.4 & 97\% & 74\% & 169\\

LLL  & 0.33 & 0.12 & 58\% & 9.4 & 94\% & 323 & 34\% & 6.9 & 100\% & 101 & 63\% & 7.1 & 3.0 & 91\% & 86\% & 188\\

LMT  & 0.28 & 0.09 & 61\% & 10.7 & 99\% & 327 & 31\% & 7.3 & 100\% & 85 & 64\% & 8.4 & 2.9 & 99\% & 84\% & 182\\

M    & 0.11 & 0.07 & 69\% & 13.9 & 100\% & 463 & 20\% & 6.3 & 99\% & 46 & 69\% & 12.2 & 2.0 & 100\% & 60\% & 238\\

MAR  & 0.15 & 0.07 & 61\% & 12.3 & 100\% & 324 & 21\% & 6.9 & 99\% & 50 & 62\% & 10.4 & 2.4 & 100\% & 71\% & 170\\

MFE  & 0.16 & 0.09 & 60\% & 10.9 & 99\% & 318 & 24\% & 7.1 & 98\% & 62 & 62\% & 8.8 & 2.5 & 99\% & 71\% & 170\\

MHP  & 0.20 & 0.10 & 62\% & 10.2 & 99\% & 377 & 25\% & 5.9 & 99\% & 62 & 64\% & 8.5 & 1.9 & 99\% & 55\% & 199\\

MHS  & 0.23 & 0.08 & 56\% & 9.2 & 99\% & 258 & 24\% & 6.6 & 100\% & 58 & 58\% & 7.7 & 2.6 & 98\% & 77\% & 139\\

MRK  & 0.10 & 0.07 & 62\% & 11.0 & 100\% & 330 & 17\% & 5.4 & 99\% & 40 & 63\% & 9.8 & 1.8 & 100\% & 55\% & 170\\

MRO  & 0.09 & 0.06 & 61\% & 11.8 & 100\% & 333 & 16\% & 6.3 & 95\% & 36 & 63\% & 10.6 & 2.0 & 100\% & 54\% & 172\\

MWV  & 0.18 & 0.10 & 62\% & 10.3 & 100\% & 330 & 28\% & 6.7 & 100\% & 75 & 64\% & 8.2 & 2.4 & 100\% & 74\% & 180\\

NEM  & 0.20 & 0.07 & 56\% & 9.9 & 99\% & 253 & 20\% & 6.1 & 99\% & 47 & 58\% & 8.6 & 2.5 & 99\% & 75\% & 135\\

OMC  & 0.15 & 0.09 & 57\% & 10.1 & 99\% & 286 & 20\% & 6.4 & 98\% & 48 & 59\% & 8.6 & 2.4 & 98\% & 73\% & 151\\

PCS  & 0.11 & 0.18 & 62\% & 7.1 & 96\% & 411 & 18\% & 3.7 & 97\% & 54 & 63\% & 6.5 & 0.7 & 93\% & 20\% & 214\\

PHM  & 0.07 & 0.08 & 64\% & 10.2 & 100\% & 384 & 15\% & 5.5 & 90\% & 34 & 65\% & 9.4 & 1.2 & 99\% & 40\% & 195\\

PKI  & 0.11 & 0.11 & 55\% & 7.8 & 99\% & 266 & 20\% & 4.8 & 97\% & 47 & 57\% & 6.7 & 1.8 & 98\% & 53\% & 141\\

R    & 0.27 & 0.11 & 56\% & 8.6 & 98\% & 259 & 28\% & 6.0 & 100\% & 74 & 59\% & 6.9 & 2.9 & 97\% & 85\% & 147\\

RAI  & 0.25 & 0.10 & 61\% & 9.2 & 99\% & 334 & 28\% & 5.7 & 100\% & 73 & 63\% & 7.6 & 2.4 & 99\% & 71\% & 182\\

SLB  & 0.24 & 0.07 & 62\% & 12.0 & 99\% & 330 & 19\% & 5.5 & 98\% & 46 & 63\% & 10.6 & 1.7 & 99\% & 51\% & 171\\

TE   & 0.09 & 1.69 & 60\% & 8.0 & 98\% & 371 & 18\% & 4.4 & 85\% & 48 & 61\% & 7.2 & 1.3 & 98\% & 39\% & 196\\

TWC  & 0.25 & 0.10 & 55\% & 9.7 & 99\% & 253 & 27\% & 6.6 & 100\% & 73 & 58\% & 7.6 & 3.0 & 99\% & 81\% & 142\\

WHR  & 0.34 & 0.11 & 56\% & 8.2 & 97\% & 272 & 29\% & 6.3 & 100\% & 78 & 59\% & 6.6 & 2.9 & 95\% & 86\% & 156\\

WIN  & 0.06 & 0.26 & 48\% & 3.9 & 79\% & 340 & 10\% & 2.8 & 50\% & 34 & 49\% & 3.7 & 0.6 & 79\% & 29\% & 179\\

WPI  & 0.22 & 0.10 & 61\% & 9.6 & 98\% & 361 & 28\% & 5.8 & 100\% & 75 & 64\% & 7.7 & 2.2 & 98\% & 71\% & 196\\

XTO  & 0.08 & 0.06 & 53\% & 10.9 & 100\% & 238 & 15\% & 6.5 & 100\% & 32 & 55\% & 9.6 & 2.7 & 100\% & 78\% & 125\\
\hline
Grand mean & 0.18 & 0.18 & 58\% & 9.8 & 98\% & 313 & 23\% & 6.0 & 97\% & 58 & 61\% & 8.3 & 2.3 & 97\% & 67\% & 168\\
\hline
\end{tabular}
\end{center}
\vskip 12pt
\scriptsize

\noindent Table 5 presents the average results of regressions:
\vskip 6pt
$|\Delta P_{k}|=\hat{\alpha}_{O,i} + \hat{\beta}_{O,i}|OFI_{k}|+\hat{\epsilon}_{O,k}$,

$|\Delta P_{k}|=\hat{\alpha}_{V,i} + \hat{\beta}_{V,i}VOL^{\hat{H}_i}_{k}+\hat{\epsilon}_{V,k}$,

$|\Delta P_{k}|=\hat{\alpha}_{W,i} + \hat{\phi}_{O,i}|OFI_{k}| + \hat{\phi}_{V,i}VOL^{\hat{H}_i}_{k} +\hat{\epsilon}_{W,k}$,
\vskip 6pt
\noindent where $\Delta P_{k}$ are the 10-second mid-price changes, $OFI_{k}$ are the contemporaneous order flow imbalances and $VOL_{k}$ are the contemporaneous trade volumes. The exponents $\hat{H}_i$ were estimated in each subsample beforehand using a logarithmic regression: $\log{|\Delta P_{k}|}= \log{\hat{\bar{\theta}}_i}+\hat{H}_i\log{VOL_{k}}+\log{\hat{\xi}_{k}}$. These regressions were estimated using 273 half-hour subsamples (indexed by $i$) for each stock and their outputs were averaged across subsamples. Each subsample typically contains about 180 observations (indexed by $k$). The t-statistics were computed using White's standard errors. For each of three regressions, Table 5 reports the average $R^2$, the average t-statistic of the coefficient(s), the percentage of samples where the coefficient(s) passed the z-sest at the 5\% significance level and the F-statistic of the regression.

\normalsize
\newpage

\addtolength{\oddsidemargin}{.5in}
\addtolength{\evensidemargin}{.5in}

\bibliographystyle{siam}
\bibliography{orderimbalance}

\newpage
\appendix

\section {Appendix: TAQ data processing}

Quotes data were filtered as follows:
\begin{enumerate}
\item Timestamp $\in$ [9:30 am, 4:00 pm].
\item Bid, ask, bid size, ask size are positive.
\item  Quote mode $\not\in \{4, 7, 9, 11, 13, 14, 15, 19, 20, 27, 28\}$
\end{enumerate}
Trades data were filtered as follows:
\begin{enumerate}
\item Timestamp $\in$ [9:30 am, 4:00 pm].
\item Price and size are positive.
\item Correction indicator $\leq 2$.
\item Condition $\not\in \{"O", "Z", "B", "T", "L", "G", "W", "J", "K"\}$
\end{enumerate}
From the filtered quotes data we construct the National Best Bid and Offer (NBBO) quotes. This is done by scanning through the filtered quotes data, while maintaining a matrix with the best quotes for every exchange. When a new entry is read, we check the exchange flag of that entry and update the corresponding row in the exchange matrix. Using this matrix, the NBBO prices are computed at each entry as the highest bid and the lowest ask across all exchanges. The NBBO sizes are simply the sums of all sizes at the NBBO bid and ask across all exchanges.

After the NBBO quotes are computed, we applied a simple quote test to the NBBO quotes and the filtered trades data. This test matches trades with NBBO quotes and computes the direction of matched trades. A trade is matched with a quote, if:

\begin{enumerate}
\item Trade is not inside the spread, i.e.
\begin{enumerate}
\item Trade price $\geq$ NBBO ask: in this case the trade is considered to be a buy trade.
\item Trade price $\leq$ NBBO bid: in this case the trade is considered to be a sell trade.
\end{enumerate}
\item Trade date $=$ quote date.
\item Trade timestamp $\in$ [quote timestamp, quote timestamp + 1 second].
\item If the above conditions allow to match a trade with several quotes, it is matched with the earliest quote.
\end{enumerate}

There are other routines to estimate trade direction, including the tick test and the Lee-Ready rule  \cite{lee91}. Although the latter is used quite frequently, there seems to be no compelling evidence of superiority of either of these heuristics \cite{odders00,theissen01}. To test the robustness of our findings to the choice of a trade direction test, we compared our results on a subsample of data, applying alternatively the tick test or the quote test and it led to virtually the same results.

Finally, we removed observations with extremely high bid-ask spreads. To apply this filter coherently across stocks, we computed for each stock the 95-th percentile of its bid-ask spread distribution and removed the 5\% of that stock's quotes with the spreads above that percentile.
\end{document}